\documentclass[useAMS,usenatbib]{mn2e}
\usepackage{graphicx,amssymb,times,multirow}

\title[Long-term evolution of three-planet systems]
{Long-term evolution of three-planet systems to the post-Main Sequence and beyond}
\author[Mustill et al.]{Alexander J. Mustill$^{1}$\thanks{E-mail: alex.mustill@uam.es},  Dimitri Veras$^{2,3}$\thanks{E-mail: veras@ast.cam.ac.uk} and Eva Villaver$^1$\thanks{E-mail: eva.villaver@uam.es}\\
$^{1}$Universidad Aut\'{o}noma de Madrid, Departamento de F\'{i}sica Te\'{o}rica, 28049 Madrid, Spain
\\
$^{2}$Department of Physics, University of Warwick, Coventry CV4 7AL\\
$^{3}$Institute of Astronomy, University of Cambridge, Madingley Road, Cambridge CB3 0HA}

\begin{document}

\date{Accepted XXX. Received XXX; in original form XXX}

\pagerange{\pageref{firstpage}--\pageref{lastpage}} \pubyear{XXXX} 

\maketitle

\label{firstpage}

\begin{abstract}
We study the stability of systems of three giant planets orbiting $3-8\mathrm{\,M}_\odot$ stars at orbital distances of $>10$\,au as the host star ages through the Main Sequence (MS) and well into the White Dwarf (WD) stage. Systems are stable on the MS if the planets are separated by more than $\sim9$ Hill radii. Most systems surviving the MS will remain stable until the WD phase, although planets scattered onto small pericentres in unstable systems can be swallowed by the expanding stellar envelope when the star ascends the giant branches. Mass loss at the end of the asymptotic giant branch triggers delayed instability in many systems, leading to instabilities typically occurring at WD cooling ages of a few 100 Myr. This instability occurs both in systems that survived the star's previous evolution unscathed, and in systems that previously underwent scattering instabilities. The outcome of such instability around WDs is overwhelmingly the ejection of one of the planets from the system, with several times more ejections occurring during the WD phase than during the MS. Furthermore, few planets are scattered close to the WD, just outside the Roche limit, where they can be tidally circularised. Hence, we predict that planets in WD systems rarely dynamically evolve to become ``hot Jupiters''. Nor does it appear that the observed frequency of metal pollution in WD atmospheres can be entirely explained by planetesimals being destabilised following instability in systems of multiple giant planets, although further work incorporating low-mass planets and planetesimals is needed.
\end{abstract}


\section{Introduction}

The study of planetary systems as their host stars evolve beyond the Main Sequence (MS) is now becoming an important area of research. The detection of the pollution of the atmospheres of White Dwarfs (WDs) by comets and asteroids has allowed the composition of extra-Solar planetesimals to be determined \citep[e.g.,][]{Zuckerman+07,Gaensicke+12,XJ13}. Binaries comprising a WD and a low-mass substellar companion provide a calibration for atmospheric models of Brown Dwarfs and giant planets \citep{Pinfield+06,Steele+09,Day-Jones+11}. The population of planetary, sub-stellar and low-mass stellar companions that survive the star's giant phases can help to constrain poorly-understood aspects of tidal interactions \citep{Rasio+96,VL09,MV12} and Common Envelope evolution \citep[][Mustill et al., submitted]{SG03,Zorotovic+10,Casewell+12,PZ12,NS13}.

Many physical processes are involved in leading planetary systems from the MS through to the star's WD stage. Planets and planetesimals see their orbital radii increased by stellar mass loss, which in extreme cases can also cause eccentricity excitation and even ejection of distant planets and comets \citep{Omarov62,Hadjidemetriou63,PA98,Veras+11}. This mass loss can also affect the system's many-body dynamics, as planet:star mass ratios increase \citep{DS02,BMW11,Debes+12,Veras+13,Voyatzis+13,Adams+13}. Planets closer to the star will feel the effects of photoevaporation \citep{VL07} and tidal forces \citep{VL09,Kunitomo+11,MV12,NS13}, while smaller bodies may be sublimated or destroyed by stellar wind drag \citep{Jura08,BW10,JX10}. Achieving a full understanding of the behaviour of planetary systems around evolving stars is therefore a complex problem.

Around 1 in 5 of all known planetary systems has more than one known planet\footnote{\texttt{exoplanet.eu, exoplanets.org}}, a fraction that may increase as more difficult-to-detect companions to existing planets are found. These systems range from compact systems such as Kepler-11 \citep{Lissauer+11}, with 6 known planets within 0.5\,au of the host star, to wider ones such as HR~8799, with four known planets beyond 10\,au \citep{Marois+10}. Many-body dynamics is therefore important for a significant fraction and wide diversity of planetary systems. Moreover, many multi-planet systems are dynamically ``packed'', \emph{i.e.,} planets cannot be much closer together, nor can more planets be introduced into gaps between them, without the systems being violently unstable \citep{BarnesQuinn04,BarnesRaymond04,RaymondBarnes05,FangMargot13}. The seminal studies of \cite{DL98,DS02} showed that mass loss can destabilise systems that are stable in the absence of mass loss, or induce instability more rapidly in unstable systems, as planet--planet interactions strengthen when the star loses mass.

\cite{DS02} studied both two- and three-planet systems; however, they did not self-consistently study the dynamical evolution of systems before, during and after mass loss. In \citet[hereafter Paper~I]{Veras+13}, we set out to rectify this, incorporating the stars' mass-loss history into N-body integrations and following the evolution of two-planet systems from the zero-age MS through to WD cooling ages of up to 5\,Gyr.

In this Paper, we extend our previous work to systems of three planets, where the stability properties are more complex and less well understood. In common with our earlier work, we focus on stellar masses of $3\mathrm{\,M}_\odot$ and above, for computational reasons, to giant planets, as these are the easiest to detect, and to planets beyond 10\,au, to avoid complicating the evolution with tides, photoevaporation, or Common Envelope evolution.

The Paper is laid out as follows. In Section\,\ref{sec:stab} we discuss semi-analytical insights into the expected stability of planetary systems at different stellar evolutionary phases. In Section\,\ref{sec:sims} we describe our N-body simulations of three-planet systems. In Section\,\ref{sec:discussion} we discuss our results. We conclude in Section\,\ref{sec:conclusions}.

\section{Stability estimates} 

\label{sec:stab}

Because the parameter space for the initial conditions of arbitrary 3-planet systems is so large, we seek some analytical or semi-analytical guidance to help to inform our choice of initial conditions for our numerical integrations. The problem we face is one of determining the stability of a given planetary system orbiting a given star, over a given time-scale, in particular those critical time-scales determined by the stellar mass such as the MS lifetime.

The stability of two-planet systems throughout the lifetimes of their $3-8\mathrm{\,M}_\odot$ host stars was studied in detail in Paper~I. In this case there is an analytical result on stability, the \emph{Hill stability criterion,} stating under what conditions a system may see a collision between the two planets \citep{Gladman93}. However, in Paper~I and \cite{VM13} we showed that the Hill criterion can underestimate instability which can occur on long time-scales comparable to a star's MS lifetime, since it does not account for Lagrange instability, which typically results in the ejection of a planet.  Moreover, the four-body problem is complex enough that often one
cannot apply two-planet stability criteria pairwise amongst the three
planets. 
 In fact, three planets which are well-separated enough so that
adjacent pairs of planets satisfy the two-body
stability criteria may eventually suffer collisions
and close encounters.  The reasons for this behavior are due
to the properties of Kolmogorov--Arnol'd--Moser tori embedded in a chaotic sea,
and are explored in detail in \cite{Shikita+10}.

As no useful analytical parallel to the Hill criterion has yet been found for three-planet systems, the most general stability results available are afforded by semi-analytical fits to N-body integrations. To reduce the multi-dimensional parameter space that three-planet systems offer, we make the commonly-used simplifications that all planets are of equal mass and initially have no orbital eccentricity. The former has no strong effect on stability times, although it does accentuate the effect of mean motion resonances \citep[MMRs][]{Chambers+96}. The latter is justified by the fact that, provided planets are sufficiently close to experience instability, the outcome of the subsequent scattering is only weakly dependent on the initial eccentricity distribution \citep{JuricTremaine08}. We also consider systems in which planets' orbits increase in geometrical progression, where the inner and outer planet pairs are separated by the same number of Hill radii as described below. In such systems, several authors have attempted to fit the stability time-scales, usually defined as the time until planets' orbits first intersect, found from N-body integrations by functions of the form
\begin{equation}\label{eq:logt-generic}
\log t = b\Delta+c,
\end{equation}
$\Delta$ being the planets' separation in Hill radii \citep{Chambers+96,DS02,FaberQuillen07,Zhou+07,SmithLissauer09}. Such functions are hard to generalise as the coefficients depend on the planet masses \citep{FaberQuillen07}. The scatter in these relations can also be large: up to two orders of magnitude in the stability time for systems of similar separations, particularly when near MMRs \citep{SmithLissauer09}. Furthermore, this dependence of $\log t$ on $\Delta$ appears to steepen at long times of $>10^8$ orbital periods \citep{SmithLissauer09}. Nevertheless, as attempts to analytically calculate these timescales are yet at an early stage \citep{Shikita+10,Quillen11}, this represents the best guide to set up our simulations.

In order to predict those three-planet systems that will be stable over a star's MS lifetime, we must assign values to the constants in Equation\,\ref{eq:logt-generic}.
We must first define the Hill radius, which we take to be the single-planet Hill radius
\begin{equation}\label{eq:rh}
r_\mathrm{H} = a_i \left(\frac{M_i}{3 M_\star}\right)^{1/3},
\end{equation} 
where $a_i$ is a planet's semi-major axis, $M_i$ its mass, and $M_\star$ the star's mass, rather than the alternative mutual Hill radius
\begin{eqnarray}
r_\mathrm{H,\,mut} &=& \frac{a_i + a_{i+1}}{2}\left(\frac{M_i + M_{i+1}}{3 M_{\star}}\right)^{1/3} \\
&\mathrm{or }&\frac{a_i + a_{i+1}}{2}\left(\frac{M_i}{3 M_{\star}}\right)^{1/3}.
\end{eqnarray}
We make this choice because a distance in single-planet Hill radii translates linearly into a spacing in au; this is not the case for the mutual Hill radius, where if two planets at $a_i$ and $a_{i+1}$ are separated by $a_{i+1}-a_i=\Delta r_\mathrm{H,\,mut}$ then $a_{i+1}\to\infty$ as $\Delta\to2\left(M_i/3M_\star\right)^{-1/3}$.

\subsection{Instability in Main-Sequence planetary systems}

Given a MS lifetime, $t_\mathrm{MS}$
we can analytically estimate the minimum mutual planet separation, $\Delta^\mathrm{(MS)}$, 
for which stability is likely. To do this, we take Equation~3 from \cite{FaberQuillen07} and explicitly include dependencies on stellar mass and the inner planet's semi-major axis $a_\mathrm{in}$:
\begin{eqnarray}
\Delta^{(\mathrm{MS})}&=&\frac{2.6}{b}\left(\frac{M_\mathrm{pl}}{\mathrm{M_J}}\right)^{-1/12}\left(\frac{M_\star}{\mathrm{M}_\odot}\right)^{1/12} \times
\nonumber
\\
&&\bigg[\log\frac{t_{\mathrm{MS}}}{\mathrm{Myr}}-\frac{1}{2}\log\frac{M_\star}{\mathrm{M}_\odot}+\log\frac{M_{\mathrm{pl}}}{\mathrm{M_J}}
\nonumber
\\
&&-\frac{3}{2}\log\frac{a_\mathrm{in}}{\mathrm{au}}+10.0+c\bigg],\label{tinst}
\end{eqnarray}
where we set $b=3.7$ and $c=1.0$ from \cite{FaberQuillen07}. $M_\mathrm{pl}$ is the planet mass common to all planets in a system, a simplification we make henceforth throughout the paper. We note that, although these values were obtained for 10-planet rather than three-planet systems, the \cite{FaberQuillen07} formula allows the planet:star mass ratio to be varied explicitly, whereas other fits to a similar functional form \citep[e.g.,][]{Chambers+96,DS02} fit each planet mass with separate constants. This equation predicts that the stability boundary for systems of three Jupiter-mass planets on the MS lies at $\Delta^\mathrm{(MS)}\approx9r_\mathrm{H}$ (Figure\,\ref{AlexMSnew}). This is much larger than the separation for Hill stability of a two-planet system, which is at around $4.6r_\mathrm{H}$ for planets on circular coplanar orbits.

\subsection{Instability in post-Main Sequence planetary systems}

\begin{figure}
\centerline{
\includegraphics[width=0.5\textwidth]{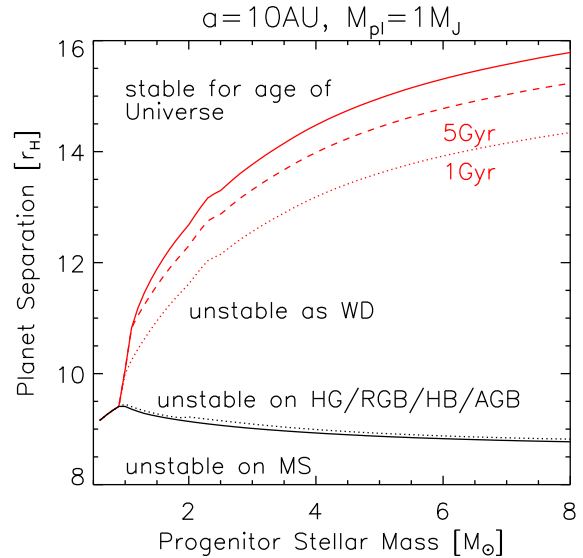} 
}
\caption{
Approximate stability behaviour of three-planet systems as a function of stellar mass and separation on the MS.
The planet mass and the innermost semi-major axis are fixed at $1\mathrm{\,M_J}$ and 1\,au, respectively.  Systems below the black solid line will be unstable 
on the Main Sequence.  Systems in the tiny strip between the black solid and
the black dotted lines will become 
unstable at some point between the end of the MS and the end of the AGB. 
Systems below the red solid line will be unstable during the WD phase, assuming that the star formed at the Big Bang; systems above this line must be stable 
for the age of the Universe. 
The red dotted and dashed lines show the stability boundaries for WD cooling ages of 1~and 5\,Gyr.  This plot demonstrates that after the star has become 
a white dwarf, many more systems can become unstable.
}
\label{AlexMSnew}
\end{figure}

When the star loses mass on the AGB, the planets' orbits expand. Assuming that the mass loss is adiabatic with respect to the planets' orbital periods\footnote{Note that mass loss adiabatic with respect to the planets' Keplerian time-scales need not imply that it is adiabatic with respect to other time-scales of interest, such as those of resonant libration or secular cycles. Non-adiabaticity with respect to these time-scales may influence aspects of the dynamics other than semi-major axis expansion.}, the semi-major axis is inversely proportional to the stellar mass: $a\propto M_\star^{-1}$. However, the planets' Hill radii expand faster than the orbits themselves due to the factor of $M_\star^{-1/3}$ in Equation~\ref{eq:rh}. This can cause previously stable systems to become unstable.  \cite{DS02} studied three-planet systems undergoing stellar mass loss. They found that including mass loss reduced the number of orbital periods required for instability at a given separation: the coefficient $b$ in Eq~\ref{eq:logt-generic} was reduced. In order to derive an explicit expression for the instability time-scale similar to Equation~\ref{tinst} for post-MS
systems, one needs to include the final stellar mass as well, and increase the planets' orbits on the assumption that they expand adiabatically. This assumption holds for intermediate separations between a few and a few hundred au \citep{VL07,Veras+11,MV12}, a condition which also ensures that any anisotropy in the mass loss is dynamically unimportant \citep*{VHT13}. Our expression for the critical separation for instability becomes:
\begin{eqnarray}
\Delta^\mathrm{(POST \ MS)}&=&\frac{2.6}{b}\left(\frac{M_\mathrm{pl}}{\mathrm{M_J}}\right)^{-1/12}\left(\frac{M_\star^\mathrm{i}}{\mathrm{M}_\odot}\right)^{1/12}\left(\frac{M_\star^\mathrm{f}}{M_\star^\mathrm{i}}\right)^{-1/4} \times
\nonumber
\\
&&\bigg[\log\frac{t_\mathrm{PMS}}{\mathrm{Myr}}-\frac{1}{2}\log\frac{M_\star^\mathrm{i}}{\mathrm{M}_\odot}+\log\frac{M_\mathrm{pl}}{\mathrm{M_J}}
\nonumber
\\
&&-\frac{3}{2}\log\frac{a}{\mathrm{AU}}+\log\frac{M_\star^{\mathrm{f}}}{M_\star^\mathrm{i}}+10.0+c\bigg]\label{eq:delta-postms}
\end{eqnarray}
Here $M_\star^\mathrm{i}$ and $M_{\star}^{\mathrm{f}}$ are the initial stellar mass and that at the end of the stage considered, while $t_\mathrm{PMS}$ is the relevant post-MS lifetime, such as a WD cooling age. $\Delta$ is measured in Hill radii of the \emph{Main Sequence} configuration, allowing us to predict the future stability of a given MS system.

\begin{figure}
  \includegraphics[width=0.5\textwidth]{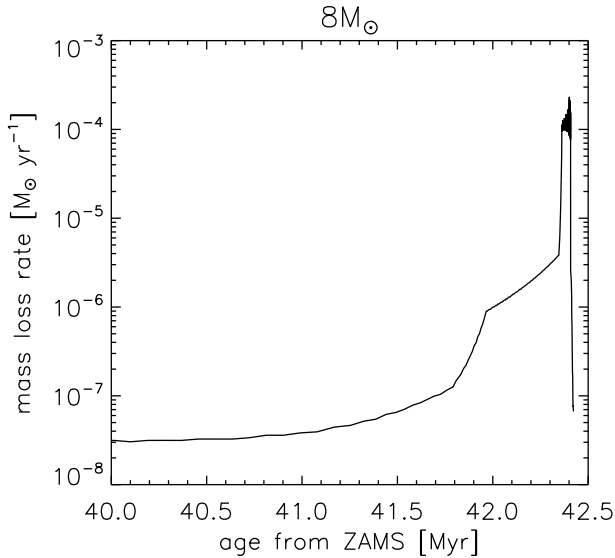}
  \caption{Mass-loss rate during the AGB for our $8\mathrm{\,M}_\odot$ stellar model.}
    \label{fig:mdot}
\end{figure}

We can now make estimates of planetary systems' stability as a function of stellar mass and evolutionary stage. We use the SSE code \citep*{Hurley+00} to find the stellar lifetimes and mass evolution of stars with metallicity $Z=0.02$ and Reimers' $\eta=0.5$. The SSE code applies Reimers mass loss at early evolutionary stages, but during the AGB it applies the semi-empirical mass-loss rate of \cite{VW93}, reaching a maximum during the superwind of $1.36\times10^{-9}(L_\star/\mathrm{L}_\odot)\mathrm{\,M}_\odot\mathrm{\,yr}^{-1}$, where $L_\star$ is the stellar luminosity. An example of the mass loss rate history on the AGB is shown in Figure~\ref{fig:mdot}, for an $8\mathrm{\,M}_\odot$ star. Mass loss from the lower-mass stars is less rapid, but still attains a maximum of $3\times10^{-5}\mathrm{\,M}_\odot\mathrm{\,yr}^{-1}$ for a $3\mathrm{\,M}_\odot$ star. The fraction of mass lost by the WD stage is 75\% for a $3\mathrm{\,M}_\odot$ star and 82\% for an $8\mathrm{\,M}_\odot$ star. 

Up to the end of the AGB, we take $t_\mathrm{PMS}$ to be the elapsed lifetime of the star.
For the WD phase, we assume that the 
system clock is reset following mass loss on the AGB, and take the relevant age to be the cooling age of the WD; this may underestimate the extent of 
instability if systems become mildly excited on the MS. 
 In Figure~\ref{AlexMSnew} we show the stability limits estimated from Equations~\ref{tinst} and~\ref{eq:delta-postms}, for the MS lifetime, the end of the AGB, WD cooling ages of 1 and 5 Gyr, and a total age of 13.7 Gyr. From the Figure, we make the following predictions:
\begin{itemize}
\item On the MS, planets with $\Delta \lesssim9 r_\mathrm{H}$ will be susceptible to instability (black solid line). This separation is remarkably insensitive to stellar mass, and hence the very different MS lifetimes, since the stellar lifetime only enters equation~\ref{tinst} logarithmically.
\item The relatively short time between the end of the MS and the end of the AGB will not allow many more systems to become unstable (black dotted line).
\item After the star has become a WD, many more systems can become unstable (red solid line), which is largely attributable to the power-law dependence of the critical separation on the final:initial 
stellar mass ratio (the increase in Hill sphere size).  More systems become 
unstable around higher-mass progenitor stars, because the mass ratio becomes more unequal. The stability boundary increases to around $16r_\mathrm{H}$ for $8\,M_\odot$ stars. Almost all of the increase is accounted for by the change in Hill radius, with the power-law dependence on $M_\star^\mathrm{f}/M_\star$ in Equation~\ref{eq:delta-postms} accounting for an increase in critical separation to $\sim14r_\mathrm{H}$ at $8\mathrm{\,M}_\odot$, the logarithmic terms making up the difference.
\item The 1~and 5\,Gyr lines (red dotted and dashed) illustrate that most WD instabilities will occur early
in the WD lifetime.
\end{itemize}

These stability limits exhibit a dependence on both planet mass and semi-major axis. As either mass or semi-major axis is reduced, the critical separation measured in Hill radii increases. This dependence is however weak: For a $5\mathrm{\,M}_\odot$ MS star, reducing the innermost planet's semi-major axis from 1000\,au to 0.1\,au, or planet mass from $10\mathrm{\,M_J}$ to $10^{-4}\mathrm{\,M_J}$, each results in an increase in the critical separation of about 4 Hill radii. We discuss this further in Section~\ref{sec:generality}.

\section{N-body Simulations}

\label{sec:sims}

\begin{figure*}
  \begin{minipage}{\textwidth}
    \includegraphics[width=.45\textwidth]{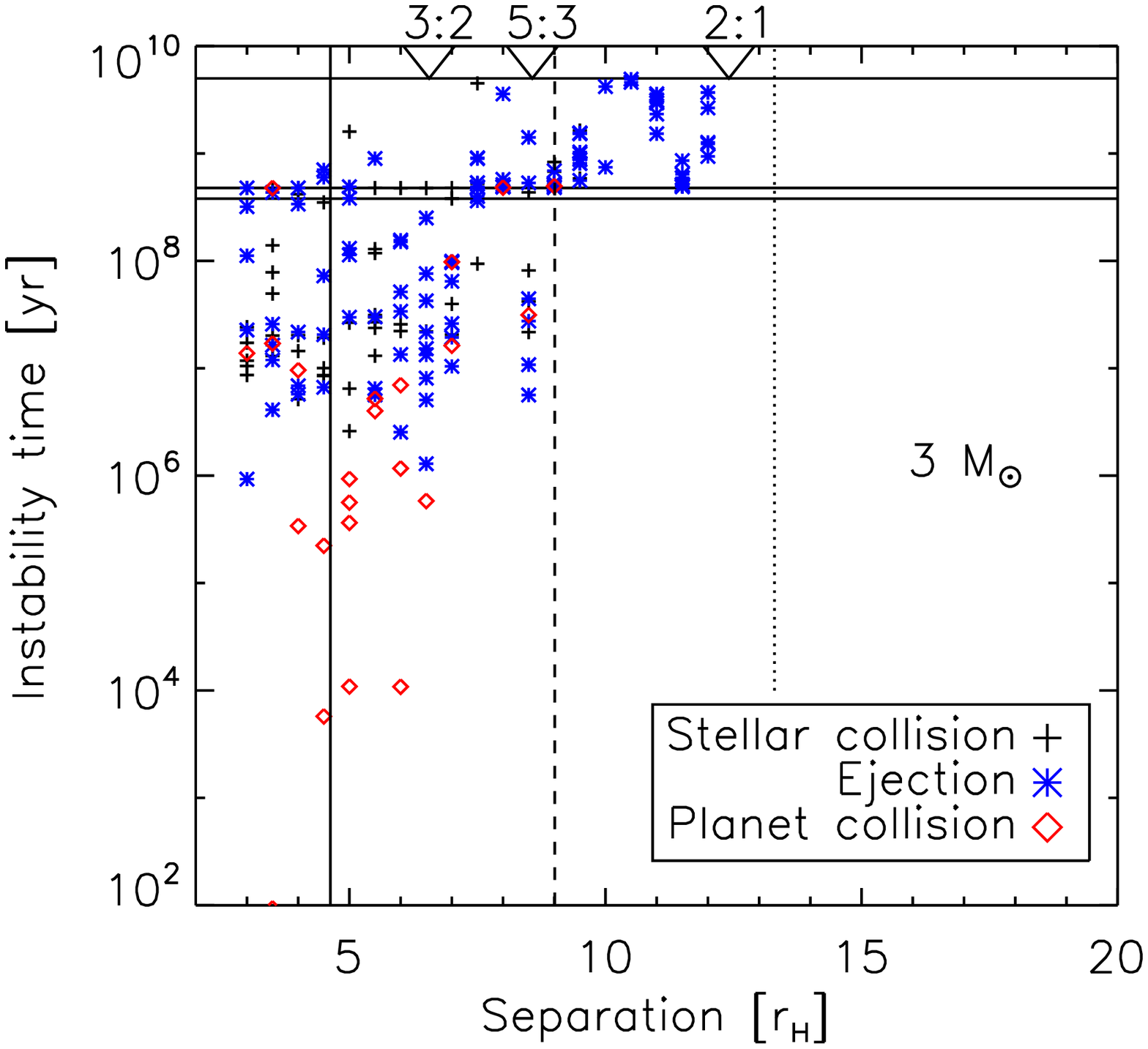}
    \includegraphics[width=.45\textwidth]{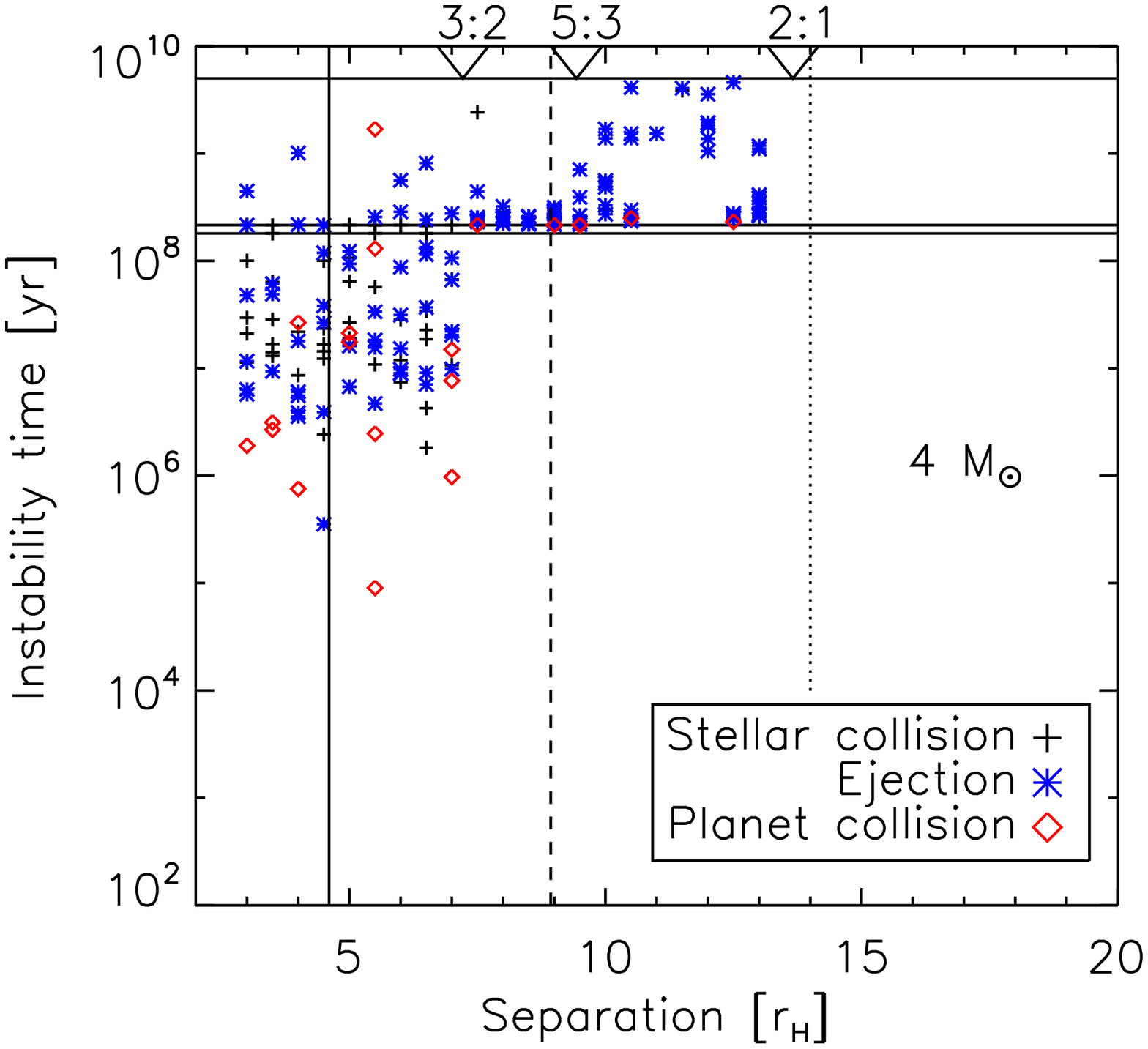}
  \end{minipage}
  \begin{minipage}{\textwidth}
    \includegraphics[width=.45\textwidth]{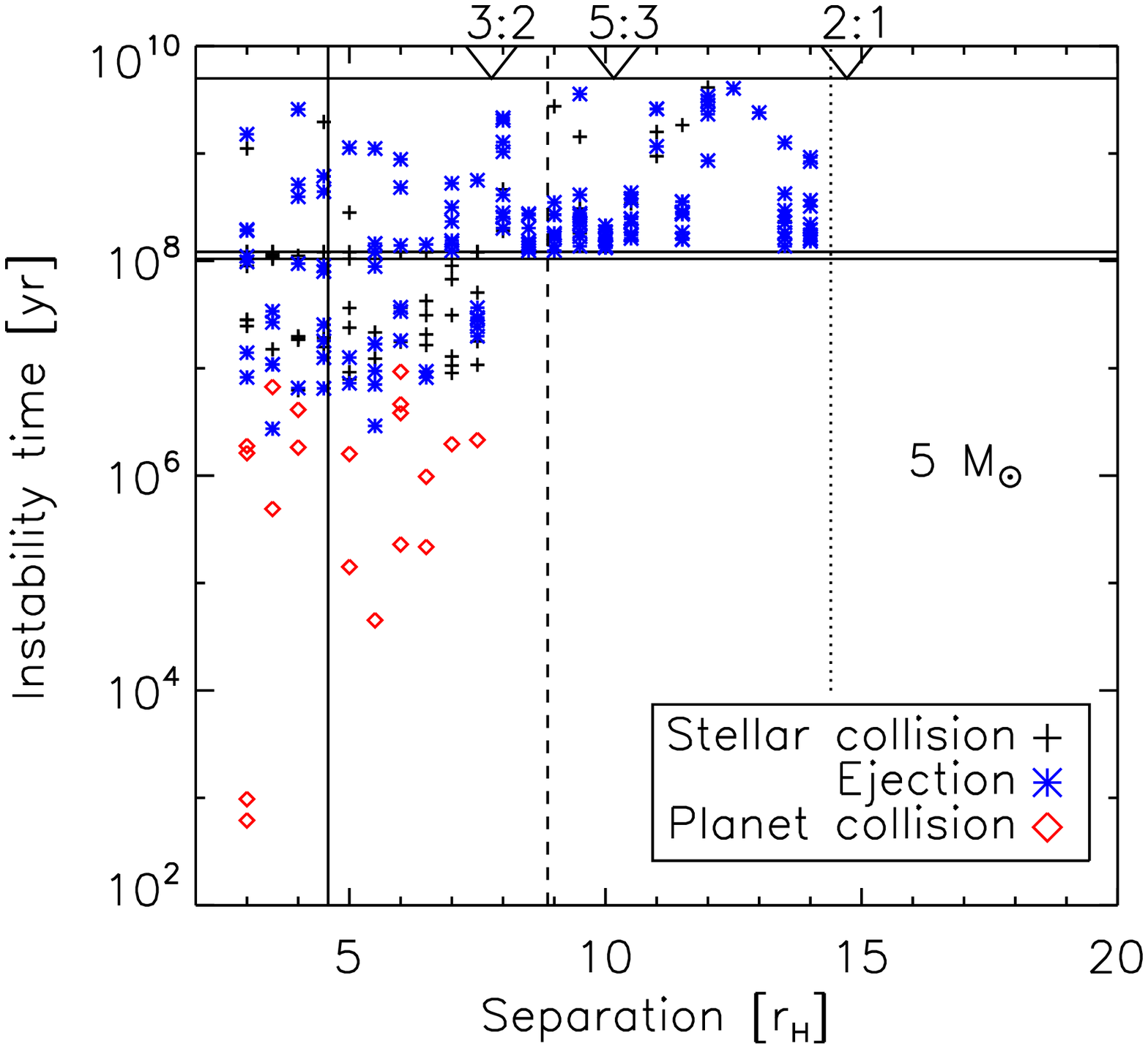}
    \includegraphics[width=.45\textwidth]{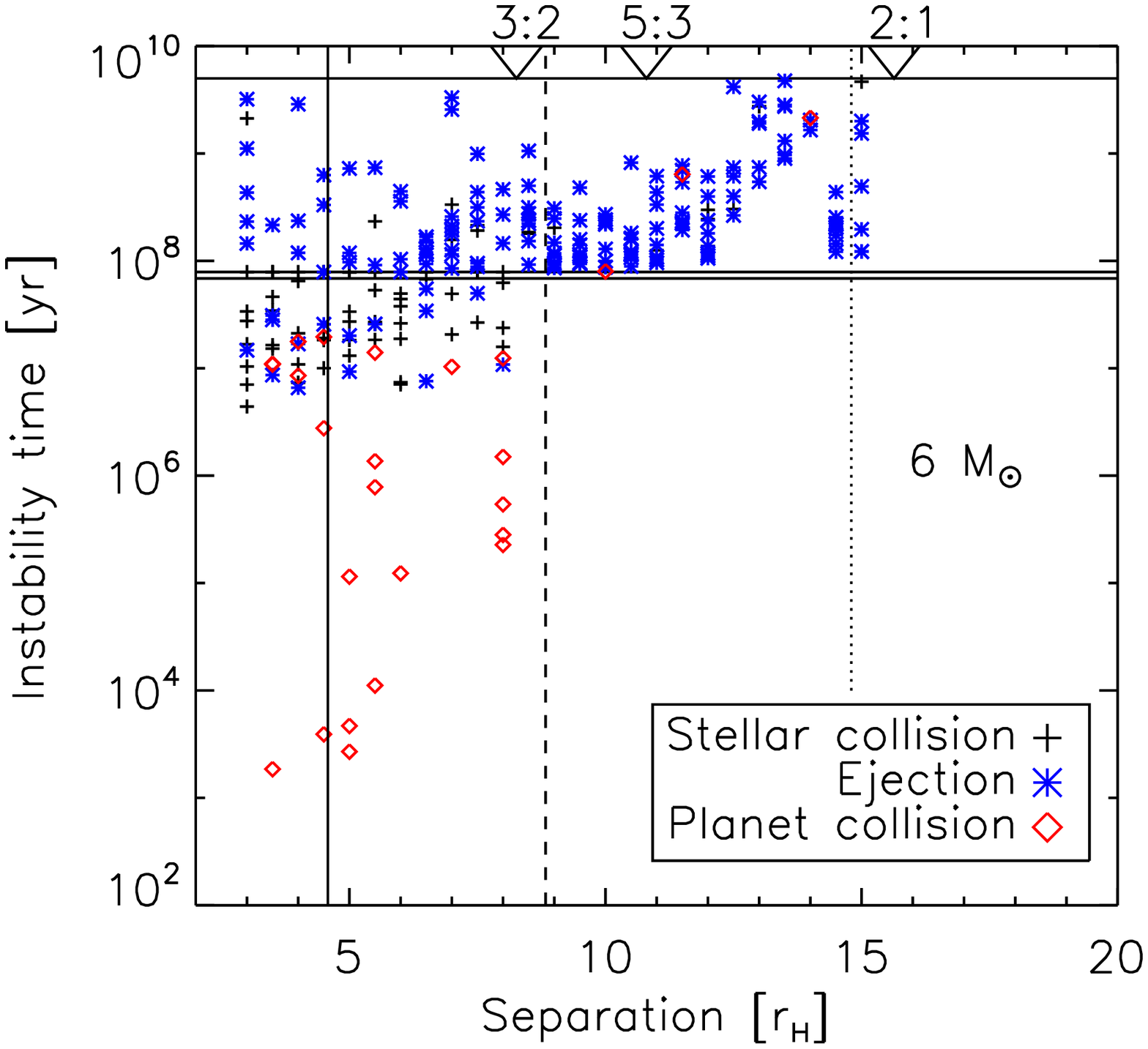}
  \end{minipage}
  \begin{minipage}{\textwidth}
    \includegraphics[width=.45\textwidth]{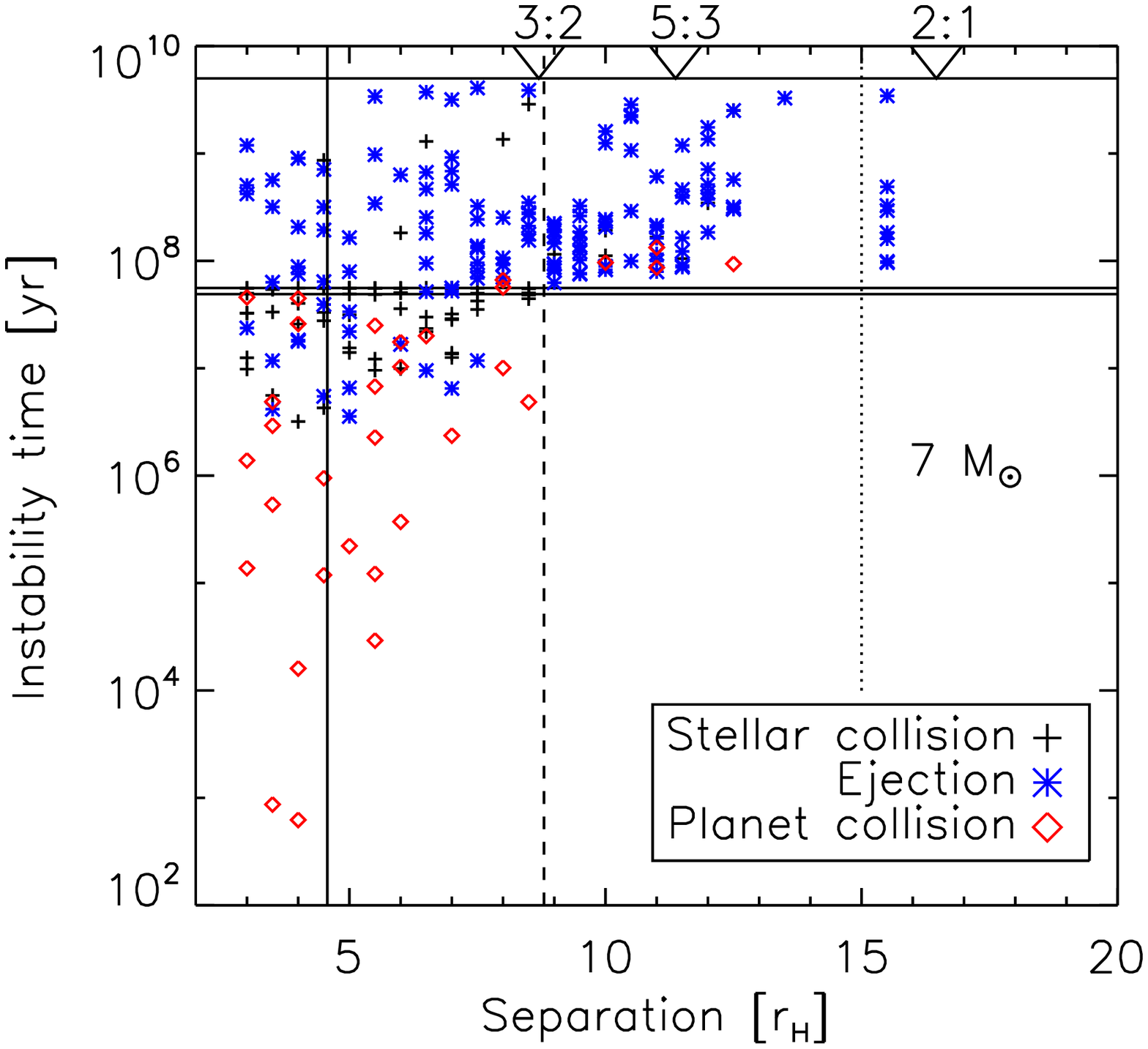}
    \includegraphics[width=.45\textwidth]{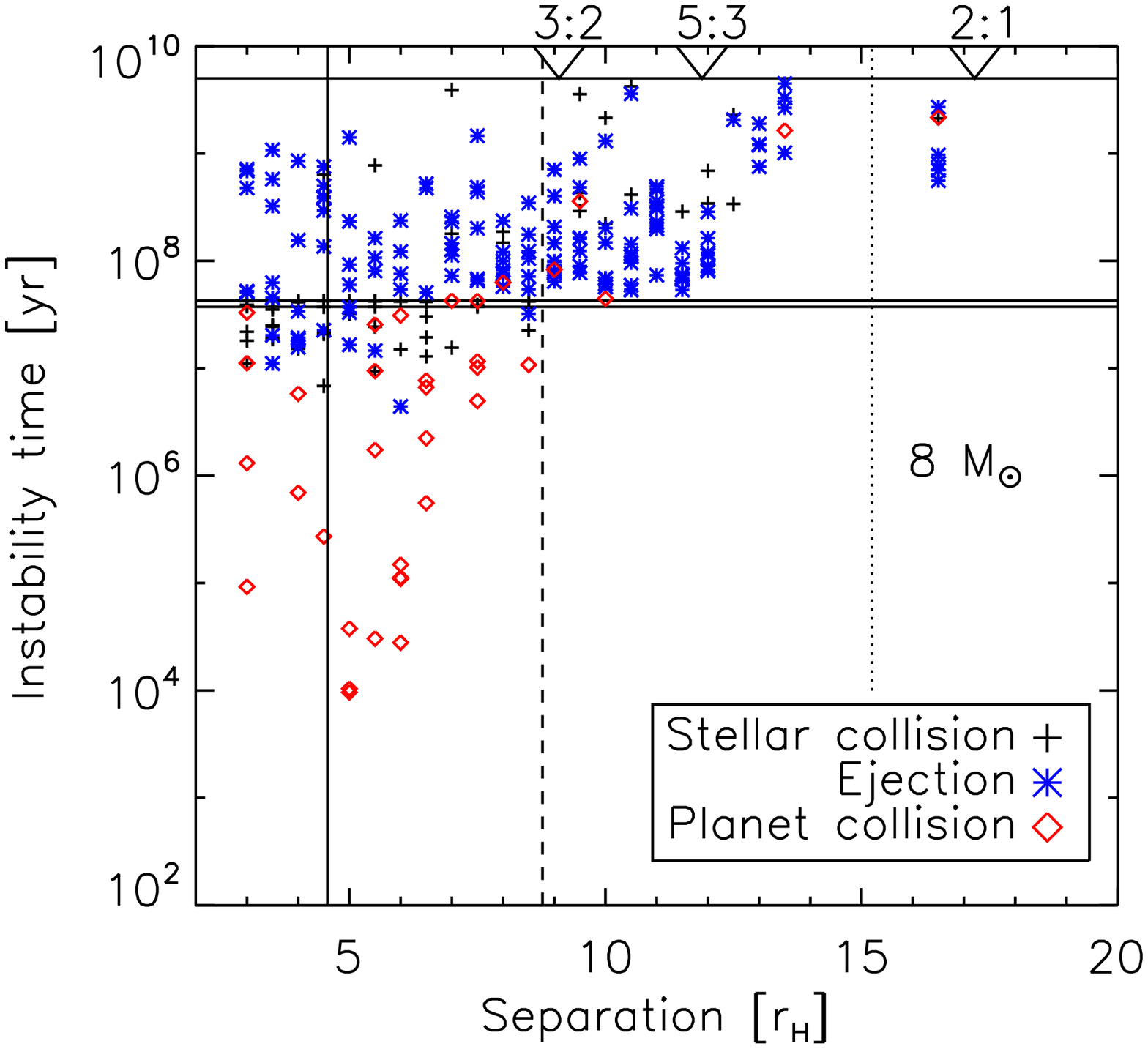}
  \end{minipage}
  \caption{Instability times in systems of three $1\mathrm{\,M_J}$ planets. Symbols are plotted every time a planet is lost; symbol types show whether planets were lost due to ejection or due to collision with another planet or the star. Horizontal lines mark the end of the MS, the beginning of the WD cooling track, and the 5\,Gyr duration of the integrations. Solid vertical lines mark the two-planet Hill stability limit on the MS, and the dashed and dotted lines mark the three-planet stability limits for MS and WD stars determined in Figure 1. Important mean motion resonances are marked with triangles.}
  \label{fig:times-by-type}
\end{figure*}

We proceed to numerically investigate the stability of systems of three equal-mass planets around stars of $3-8\mathrm{\,M_\odot}$. As discussed in Paper I, the MS lifetimes of these stars are sufficiently short to permit a statistically meaningful number of integrations to be conducted over the whole stellar lifetime and well into the WD phase. We couple the SSE stellar models from \cite{Hurley+00} to the \textsc{Mercury} Bulirsch--Stoer integrator \citep{Chambers99} as described in Paper I. We make an additional change in addition to those described in said paper, in that we replace the stellar radius with the Roche limit for tidal destruction given by
\begin{equation}
a_\mathrm{Roche}=\left(3\rho_\star/\rho_\mathrm{pl}\right)^{1/3}R_\star,
\end{equation}
(where $\rho_\star$ and $\rho_\mathrm{pl}$ are the stellar and planetary densities and $R_\star$ is the stellar radius) in the event that the Roche limit exceeds the stellar radius, which is the case for all our WDs. This change allows us to make a more realistic estimate of the efficacy of contamination of WDs by tidally disrupted planets.

Our choice for the inner planet's orbital radius should ensure that its orbit will be unaffected by tidal forces on either the RGB or AGB. Around lower-mass stars, \cite{VL09,MV12,NS13} showed that giant planets initially at 10\,au will not feel significant tidal forces. Note however that the fate of planets around the more massive AGB stars ($>5\mathrm{\,M}_\odot$) has not been studied; extrapolating the results of \cite{MV12} and \cite{NS13} to higher masses suggests that Jovian planets at 10\,au may feel the star's tidal forces on the AGB. Also, if scattering brings planets onto orbits with smaller pericentres, they will feel tidal forces. In this Paper we ignore any tidal effects, leaving the coupling of tidal and N-body dynamics for future work.

We run two sets of simulations, one with $1\mathrm{\,M_J}$ planets and one with $10\mathrm{\,M_J}$ planets. The planets' densities used to calculate the Roche limit for tidal disruption are $1.3$ and $13\mathrm{\,g\,cm}^{-3}$ respectively. In all cases, the inner planet is initially placed at 10\,au. The second planet is placed at a separation $\Delta={3,3.5,4,\ldots,17.5,18}$ in units of the inner planet's Hill radius, a spacing that corresponds to semi-major axis ratios of $1.10-1.85$ for the $1\mathrm{\,M_J}$ planets, depending on stellar mass, and $1.31-2.84$ for the $10\mathrm{\,M_J}$ planets. The third planet is then placed at the same semi-major axis ratio from the second planet. As discussed above, at a separation of 18 Hill radii, systems should be stable throughout the WD lifetime. Our closest separation of 3 Hill's radii is well within the MS stability limit, and very closely-spaced systems may even be destabilised while the gas disc is still present \citep{LMN13}; however, we choose to go so close to see whether mass loss on the AGB can trigger a second round of instabilities in systems than have already relaxed during their MS lifetime. We chose the closest separation \emph{post hoc} by finding the first separation for which all systems integrated were unstable on the MS. Planets are initially on circular orbits, measured in Jacobi co-ordinates, and are assigned small inclinations of up to $1^\circ$ in order to avoid unrealistic numbers of planet--planet collisions. We integrate each system up to total age of $5\,$Gyr, encompassing the entire MS evolution and several Gyr of WD cooling. Planets' orbital elements are recorded every 1\,Myr, and each instance of a planet being lost from a system is recorded, where the loss can be due to ejection beyond $10^4$\,au, or collision with another planet or the star. We use the loss of one or more planets as a proxy for saying a system is unstable.

In total, with six values of $M_\star$, 31 values of $\Delta$, and 8 systems with randomly chosen orbital phases at each separation, there are 1488 systems for our $1\mathrm{\,M_J}$ simulations. For the $10\mathrm{\,M_J}$ simulations we only considered the lowest stellar mass, with 248 systems integrated. A sample size of order 1000 allows the distribution of final orbital parameters to be characterised with good accuracy \citep{Chatterjee+08}.

We now describe in detail the outcomes of our $1\mathrm{\,M_J}$ integrations (Section~\ref{sec:1mj}), then the outcomes of the $10\mathrm{\,M_J}$ integrations (Section~\ref{sec:10mj}).

\subsection{Fiducial case: $M_\mathrm{pl}=1\mathrm{M_J}$}

\label{sec:1mj}

\begin{figure}
  \includegraphics[width=.5\textwidth]{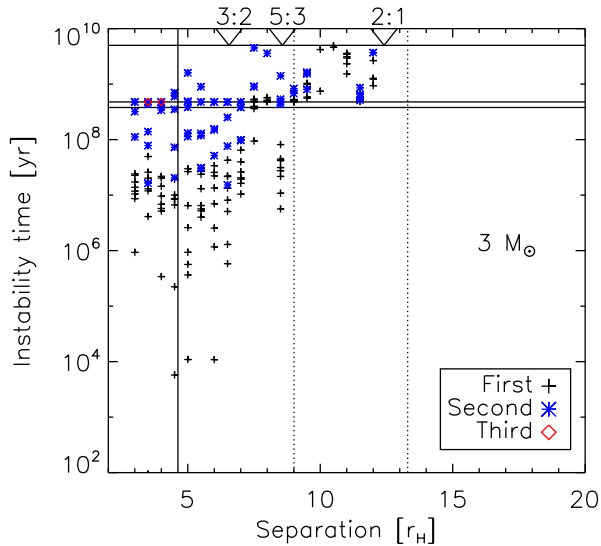}
  \caption{Time of instability as a function of planet separation, showing whether each planet lost is the first, second or third to be lost from the system. Ancillary markings are as in Figure~\ref{fig:times-by-type}.}
  \label{fig:times-by-order}
\end{figure}

\begin{figure*}
  \begin{minipage}{\textwidth}
    \includegraphics[width=0.5\textwidth]{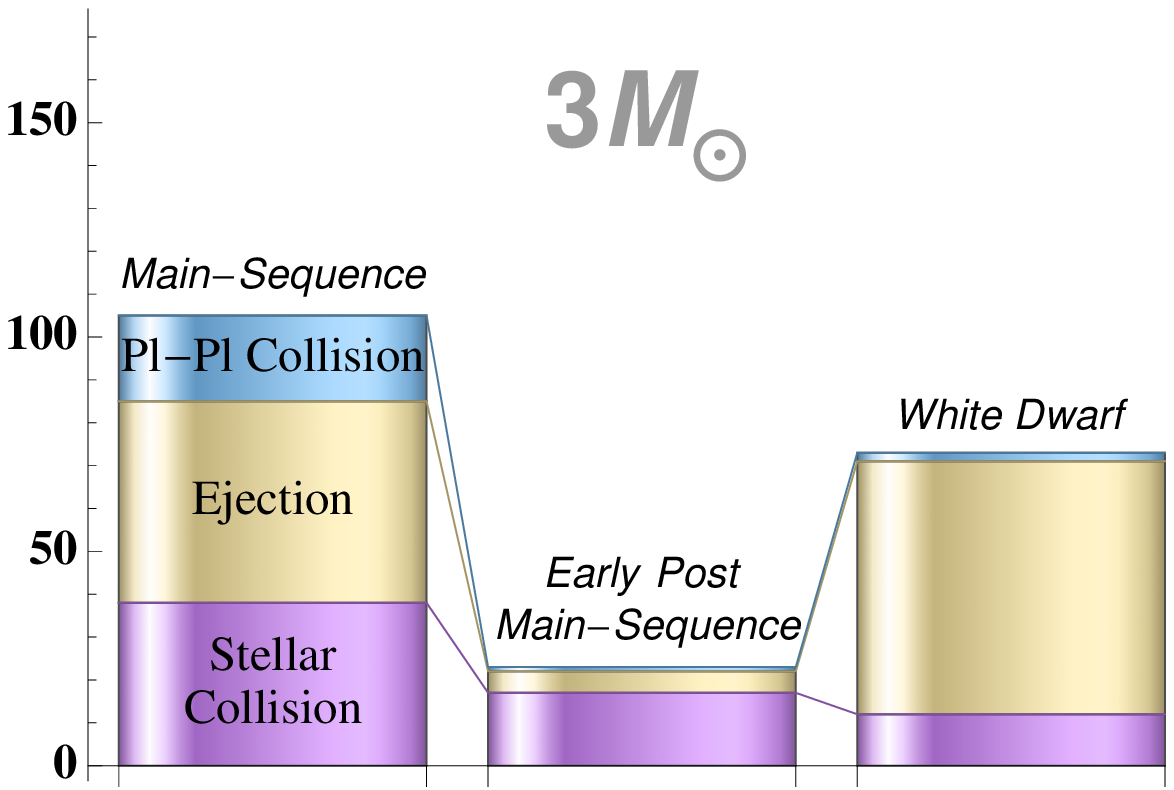}
    \includegraphics[width=0.5\textwidth]{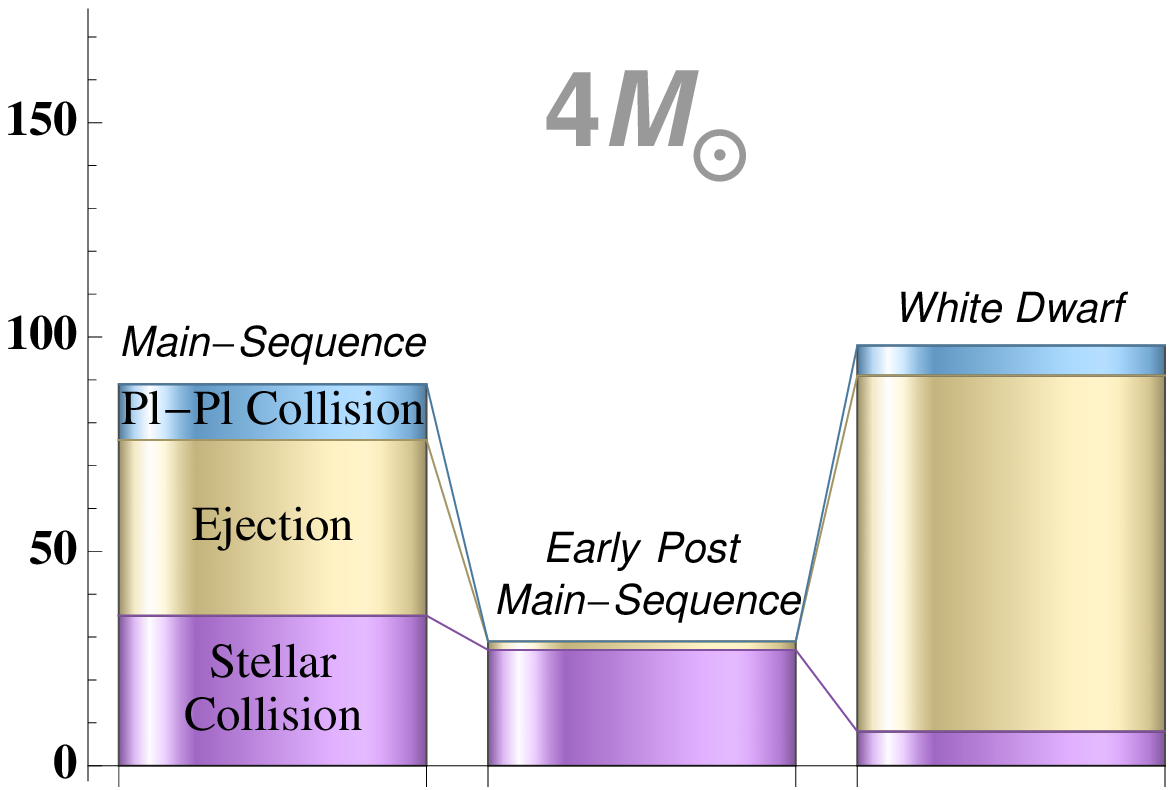}
  \end{minipage}
  \begin{minipage}{\textwidth}
    \includegraphics[width=0.5\textwidth]{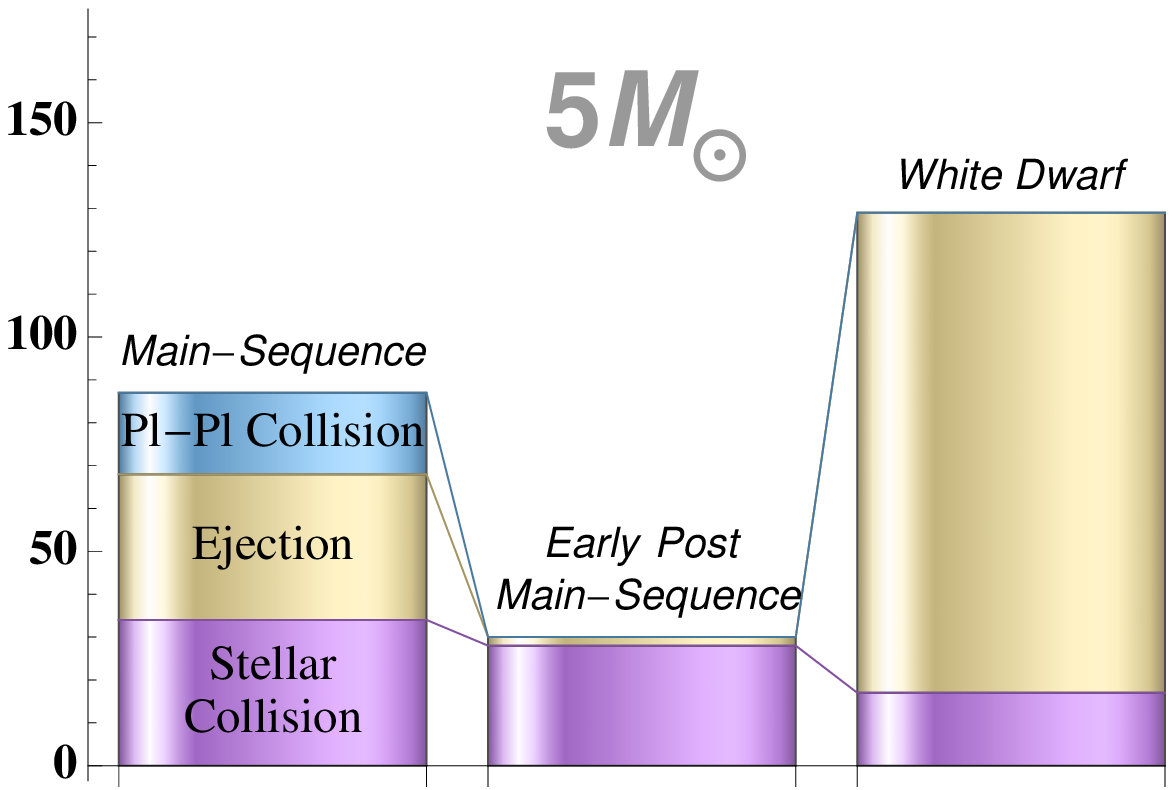}
    \includegraphics[width=0.5\textwidth]{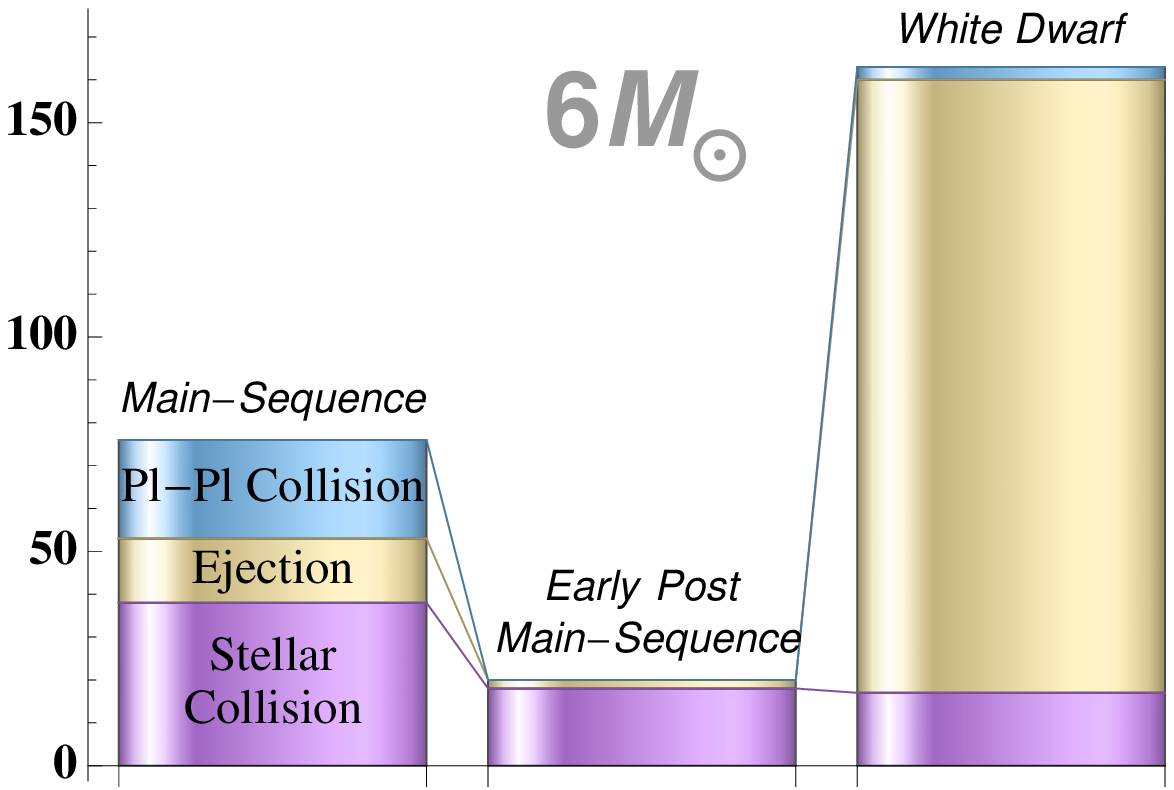}
  \end{minipage}
  \begin{minipage}{\textwidth}
    \includegraphics[width=0.5\textwidth]{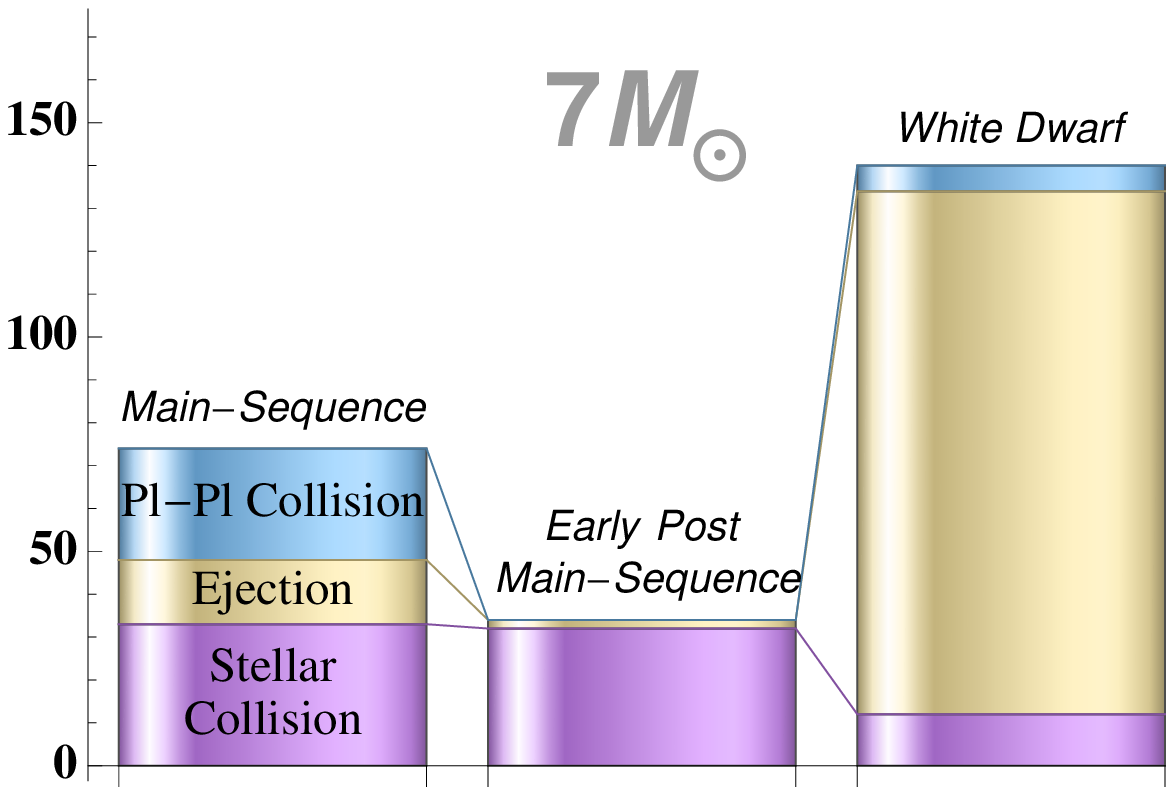}
    \includegraphics[width=0.5\textwidth]{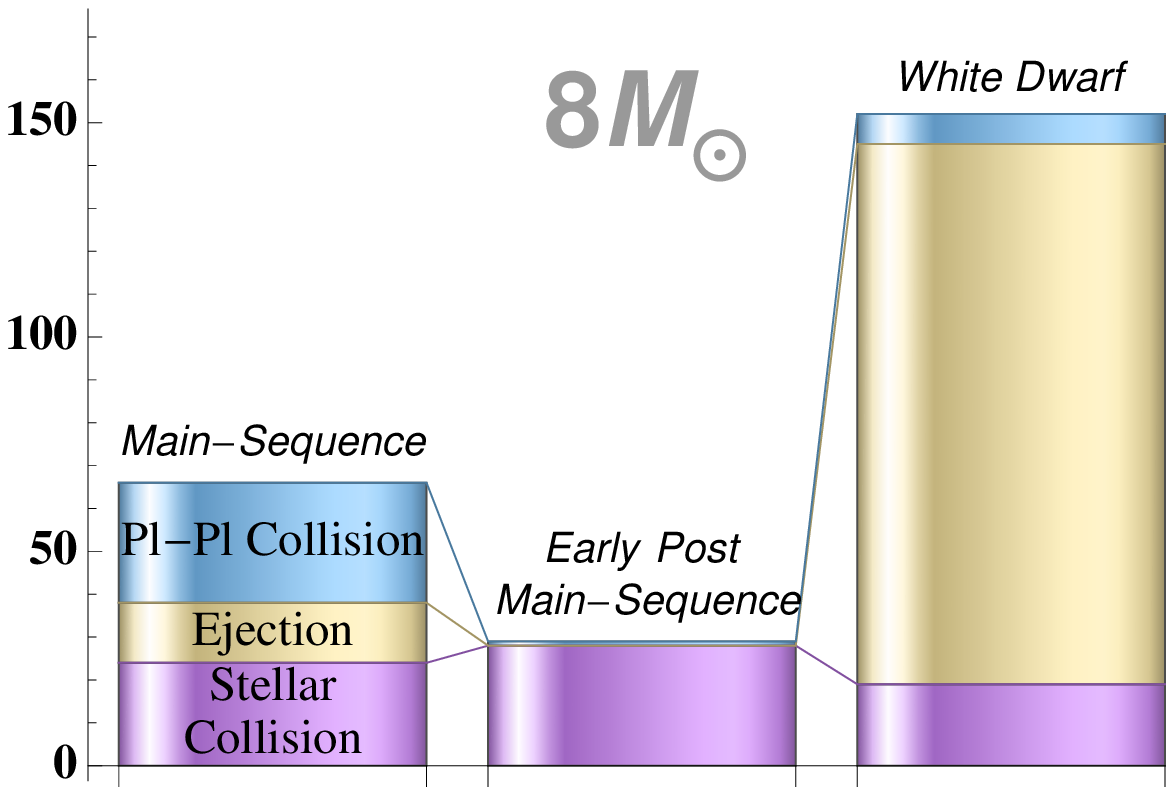}
  \end{minipage}
  \caption{Number of planets lost in the three-$1\mathrm{\,M_J}$ runs, broken down by initial stellar mass, type of loss and stellar evolutionary phase.}
  \label{fig:outcomes}
\end{figure*}

First we discuss the instabilities in our $1\mathrm{\,M_J}$ systems. The times at which a planet is lost, as well as the nature of the loss---be it through collision with the star or another planet, or ejection from the system---are plotted in Figure~\ref{fig:times-by-type} as a function of initial separation for each stellar mass. We also mark on this plot the critical separations from the two-planet Hill stability criterion on the MS (vertical solid line), as well as the estimates for three-planet stability limits on the MS and WD phases (vertical dashed and dotted lines). Important time-scales are marked on the figure with horizontal lines: the end of the MS lifetime of the star, the formation of the WD, and the 5\,Gyr duration of our integrations. We treat stellar evolution in three steps: first the MS; then the post-MS up to the formation of the WD, during which interval the stellar radius becomes very large and the star loses most of its mass; and finally the WD stage. A breakdown of the numbers of planets lost, by type of loss, stellar evolutionary stage, and stellar mass, is given in Table~\ref{tab:outcomes} and displayed graphically in Figure~\ref{fig:outcomes}.

Crudely, the outcomes of instability as a function of stellar age can be summarised as follows. The systems around MS stars see planets lost due to collisions with the star and between planets, as well as ejections; all these loss mechanisms contribute in roughly equal measure. Around post-MS stars, from the subgiant to the AGB, planets are lost overwhelmingly through collision with the expanded star, with very few ejections or planet--planet collisions. Around WDs, planets are lost primarily through ejections, with few being lost due to collisions. We now describe our simulations in more detail, going through each stage of stellar evolution.

\begin{table}
\begin{center}
\begin{tabular}{llccccc}
initial $M_\star$     & outcome & MS  & early-PMS & WD  & Total\\\hline
$3\mathrm{\,M}_\odot$ & SC      & 38  & 17        & 12  & 67\\
                      & EJ      & 47  & 5         & 59  & 111\\
                      & PC      & 20  & 1         & 2   & 23\\
                      & Total   & 105 & 23        & 73  & 201\\\hline
$4\mathrm{\,M}_\odot$ & SC      & 35  & 27        & 8   & 70\\
                      & EJ      & 41  & 2         & 83  & 126\\
                      & PC      & 13  & 0         & 7   & 20\\
                      & Total   & 89  & 29        & 98  & 216\\\hline
$5\mathrm{\,M}_\odot$ & SC      & 34  & 28        & 17  & 79\\
                      & EJ      & 34  & 2         & 112 & 148\\
                      & PC      & 19  & 0         & 0   & 19\\
                      & Total   & 87  & 30        & 129 & 246\\\hline
$6\mathrm{\,M}_\odot$ & SC      & 38  & 18        & 17  & 73\\
                      & EJ      & 15  & 2         & 143 & 160\\
                      & PC      & 23  & 0         & 3   & 26\\
                      & Total   & 76  & 20        & 163 & 259\\\hline
$7\mathrm{\,M}_\odot$ & SC      & 33  & 32        & 12  & 77\\
                      & EJ      & 15  & 2         & 122 & 139\\
                      & PC      & 26  & 0         & 6   & 32\\
                      & Total   & 74  & 34        & 140 & 248\\\hline
$8\mathrm{\,M}_\odot$ & SC      & 24  & 28        & 19  & 71\\
                      & EJ      & 14  & 0         & 126 & 140\\
                      & PC      & 28  & 1         & 7   & 36\\
                      & Total   & 66  & 29        & 152 & 247
\end{tabular}
\caption{Number of planets lost in the three-$1\mathrm{\,M_J}$ runs, broken down by initial stellar mass, type of loss (``SC'' = stellar collision, ``EJ'' = ejection, ``PC'' = planet--planet collision), and stellar evolutionary phase (``MS'' = Main Sequence, ``early-PMS'' = subgiant through to the end of the AGB, ``WD'' = White Dwarf). In all there are 1488 simulations containing 4464 planets, of which 1417 were lost.}
\label{tab:outcomes}
\end{center}
\end{table}

\subsubsection{The Main Sequence}

Instability on the MS in our integrations occurs for systems separated by up to $9r_\mathrm{H}$, as predicted by Equation~\ref{tinst}. At all these separations, planets are lost through ejection and collision with the star or another planet. The two-planet Hill stability criterion predicts that planetary collisions cannot occur at separations beyond $\sim4.6r_\mathrm{H}$, and hence fails to accurately describe the stability properties of these three-planet systems.

We also see from the integrations that, at early times on the MS, instability is dominated by planet--planet collisions, while at later times collisions with the star and ejections become more important. The number of planets lost on the MS decreases with stellar mass, from 105 for the $3\mathrm{\,M}_\odot$ case to 66 for the $8\mathrm{\,M}_\odot$ case, as a consequence of the shorter MS lifetime. Meanwhile, the fraction of ejections\footnote{Where we quote percentages in this paper, the percentage refers to the raw fraction of events recorded in our simulations, while the error bars are the $68.2\%$ confidence interval for the posterior probability density function for the underlying frequency, assuming a uniform prior \citep[][Chapter~6]{Jaynes03}. In the case of no events occurring, we give a 95\% upper limit.} falls from $44.8^{+4.9}_{-4.7}\%$ of planets lost to $21.2^{+5.9}_{-4.2}\%$, for the $3\mathrm{\,M}_\odot$ and $8\mathrm{\,M}_\odot$ stars respectively, due to the star's deeper potential well.

Systems may lose more than one planet on the MS. Figure~\ref{fig:times-by-order} shows whether each planet lost was the first, second or third from that system, for the $3\mathrm{\,M}_\odot$ star. Often, the loss of the second planet occurs considerably later than the loss of the first. Considering all stellar masses, of the 437 systems losing a planet on the MS, a further 60 ($13.7^{+1.8}_{-1.5}\%$) went on to lose a second planet on the MS as well. We note that in no cases did a star lose a third planet on the MS ($<0.2\%$); i.e., scattering events never resulted in the destruction or ejection of all of a system's planets.

\begin{table*}
\begin{center}
\begin{tabular}{lcc}
Event                                        & Fraction & Percentage\\\hline
Planets ejected on MS/number planets lost on MS [$3\mathrm{\,M}_\odot$] & 47/105 & $44.8^{+4.9}_{-4.7}$\\
Planets ejected on MS/number planets lost on MS [$8\mathrm{\,M}_\odot$] & 14/66  & $21.2^{+5.9}_{-4.2}$\\
Systems losing at least 2 planets on MS/systems losing at least 1 planet on MS & 60/437 & $13.7^{+1.8}_{-1.5}$\\
Planets colliding with star on early PMS/number planets lost on early PMS & 150/165 & $90.9^{+1.8}_{-2.7}$\\
Planets ejected on early PMS/number planets lost on early PMS & 13/165 & $7.9^{+2.6}_{-1.6}$\\
Planets colliding with another planet on early PMS/number planets lost on early PMS & 2/165 & $1.2^{+1.6}_{-0.4}$\\
Second planet lost, on early PMS/number planets lost on early PMS & 129/165 & $78.2^{+2.8}_{-3.5}$\\
Third planet lost, on early PMS/number planets lost on early PMS & 9/165 & $5.4^{+2.4}_{-1.2}$\\
Planets colliding with star on WD/number planets lost on WD & 85/755 & $11.3^{+1.3}_{-1.1}$\\
Planets ejected on WD/number planets lost on WD & 645/755 & $85.4^{+1.2}_{-1.4}$\\
Planets colliding with another planet on WD/number planets lost on WD & 25/755 & $3.3^{+1.4}_{-0.8}$\\
Two-planet systems at onset of WD that are unstable/all two-planet systems at onset of WD & 149/275 & $54.2\pm3.0$\\
Three-planet systems at onset of WD that are unstable/all three-planet systems at onset of WD & 496/1024 & $48.4\pm1.6$\\
Two-planet systems at onset of WD that are unstable/all two-planet systems at onset of WD [$3\mathrm{\,M}_\odot$] & 8/40 & $20.0^{+7.7}_{-4.9}$\\
Three-planet systems at onset of WD that are unstable/all three-planet systems at onset of WD [$3\mathrm{\,M}_\odot$] & 52/165 & $31.5^{+3.8}_{-3.4}$\\
\hline
Planets ejected on MS/number planets lost on MS & 62/81 & $76.5^{+4.0}_{-5.3}$\\
Planets ejected on WD/number planets lost on WD & 77/82 & $93.9^{+1.7}_{-3.8}$\\
Planets colliding with star on early PMS/number planets lost on early PMS & 19/26 & $73.1^{+6.8}_{-10.2}$\\
Two-planet systems at onset of WD that are unstable/all two-planet systems at onset of WD & 4/31 & $12.9^{+8.4}_{-3.9}$\\
Three-planet systems at onset of WD that are unstable/all three-planet systems at onset of WD & 61/182 & $33.5^{+3.7}_{-3.3}$
\end{tabular}
\caption{Statistical errors on fractions quoted in text. Figures above the line refer to the $1\mathrm{\,M_J}$ systems, below the line, to the $10\mathrm{\,M_J}$ systems.}
\label{tab:stats}
\end{center}
\end{table*}

\subsubsection{Early post-MS evolution}

After the end of the MS, the stellar radius increases to much larger values, with the $8\mathrm{\,M}_\odot$ star attaining 1\,au on the RGB and 8\,au at the tip of the AGB. The consequence of this is a large increase in the number of planets being engulfed by the star: between the end of the MS and the end of the AGB, an additional 150 planets were lost due to stellar collision ($90.9^{+1.8}_{-2.7}\%$ of all losses at this stage), compared to 13 more ejections ($7.9^{+2.6}_{-1.6}\%$) and two planet--planet collisions ($1.2^{+1.6}_{-0.4}\%$). In contrast to planet loss on the MS, planets lost during these stages were primarily from systems that had already lost a planet: 129 of the planets lost ($78.2^{+2.8}_{-3.5}\%$) were the system's second, and 9 ($5.4^{+2.4}_{-1.2}\%$) were the system's third. Destabilisation of new systems is rare because the short duration of the post-MS phases means that there is not much time for hitherto stable systems to experience an instability, as predicted from Equation~\ref{eq:delta-postms} and Figure~\ref{AlexMSnew}. A system may lose all its planets either when the sole survivor has a small pericentre that results in its engulfment by the expanding stellar envelope, or when it has a large apocentre which makes the stellar mass loss non-adiabatic and results in a ``Great Escape'' scenario \citep{Veras+11}. The latter occurs infrequently in our integrations however, with only four of the instances of third-planet loss being through this mechanism.

Mass loss on the AGB does not typically result in immediate instability and loss of a planet: the results of AGB mass loss are played out during the host star's WD lifetime, as described next.

\subsubsection{The white dwarf stage}

The WD stage sees by far the largest number of planets lost in our simulations, with a total of 755 planets lost. The number of planets lost on the WD exceeds the number lost on the MS in all but the $3\mathrm{\,M}_\odot$ case. These losses are overwhelmingly ejections (645, or $85.4^{+1.2}_{-1.4}\%$), with a smaller number (85, or $11.3^{+1.3}_{-1.1}\%$) of collisions with the star (that is, since the stellar radius here is smaller than the Roche limit, passages within the Roche limit), and a very few planet--planet collisions (25, or $3.3^{+0.8}_{-0.5}\%$). The greatly reduced number of planet--planet collisions compared to the MS can be attributed to the larger Safronov number $\Theta$, given by
\begin{equation}
\Theta=\frac{1}{2}\left(\frac{V_\mathrm{esc}}{V_\mathrm{orb}}\right)^2=\frac{a}{r_\mathrm{pl}}\frac{M_\mathrm{pl}}{M_\star},
\end{equation}
where $V_\mathrm{esc}$ is the escape velocity from the surface of the planet and $V_\mathrm{orb}$ its orbital velocity. This measures the effectiveness of the planet at scattering other bodies. The Safronov number increases as a result of the loss of stellar mass, which has both a direct effect on the mass term and an indirect effect via adiabatic orbit expansion, meaning that the Safronov number for the planets around WDs is a factor $\left(M_\star^\mathrm{i}/M_\star^\mathrm{f}\right)^2$ larger than for the same planet when orbiting its MS progenitor. For our MS stars, the Safronov number of the innermost planet ranges from 2.5 at $8\mathrm{\,M}_\odot$ to 6.7 at $3\mathrm{\,M}_\odot$, while for the descendant WD systems the inner planet's Safronov number ranges from 77 at $8\mathrm{\,M}_\odot$ to 107 at $3\mathrm{\,M}_\odot$.

The range of separations vulnerable to instability on the WD stage obtained from the numerical integrations is broadly in line with that predicted from Equation~\ref{eq:delta-postms} and Figure~\ref{AlexMSnew}, which is shown as the right-hand dotted vertical line in Figures~\ref{fig:times-by-type} and~\ref{fig:times-by-order}. However, there is variation about this value. In particular, particles just interior to the 2:1 resonance have noticeably shorter lifetimes than their neighbours.

The stellar mass loss just before the formation of the WD causes both instability in previously stable systems, and renewed instability in systems that had previously lost a planet. Instabilities occur at about the same rate in systems that have and have not already seen an instability before the end of their AGB evolution: of 1024 three-planet systems surviving to the formation of the WD, 496 ($48.4\pm1.6\%$) experienced a subsequent instability, while of 275 two-planet systems at the formation of the WD, 149 ($54.2\pm3.0\%$) experienced a subsequent instability.

\begin{figure}
  \includegraphics[width=.5\textwidth]{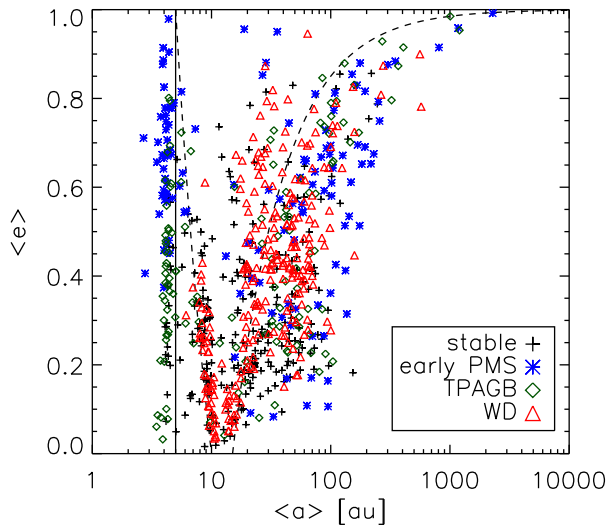}
  \caption{For systems which lose precisely one planet on the MS, the average $a$ and $e$ of the two surviving planets for the remainder of their MS lifetime is shown. Symbols show their subsequent fate: whether they lose a second planet during early post-MS evolution before the TPAGB, during the TPAGB, or the WD stage, or remain stable. The vertical solid line shows a semi-major axis of 5\,au. Dashed lines show an apocentre of 10\,au and a pericentre of 15\,au. All stellar masses are considered, with planets of $1\mathrm{\,M_J}$.}
  \label{fig:a-e-ms}
\end{figure}

However, the outcome of early instabilities can play a role in the systems' future evolution. The effects of MS instabilities on the subsequent fates of planetary systems is shown in Figure~\ref{fig:a-e-ms}. For systems which lose one planet during the MS, the average $a$ and $e$ of their surviving planets for the remainder of the MS are shown. The subsequent fate of the systems is also shown: whether they lose a second planet during the subgiant through to early AGB phases, during the TPAGB phase, or as a WD, or whether they subsequently remain stable. The orbital elements of surviving planets following the first instability clearly affects the subsequent evolution: during the post-MS stages, systems with one planet on an orbit with a small semi-major axis lose it due to engulfment, as the stellar radius expands, the higher eccentricity systems being swallowed first when the stellar radius is smaller. During the WD phase, the planets lost come from more closely-packed systems with a higher $e$ for a given $a$ on the MS.

The distribution of $a$ and $e$ in Figure~\ref{fig:a-e-ms} is similar to that seen in other scattering studies \citep[e.g.,][]{Chatterjee+08}. A large population of planets ends up at $\sim4$\,au. This implies a change in orbital energy, relative to an initial orbit at 10\,au, just greater than that required to unbind a co-orbital planet, which would move a surviving planet in to 5\,au. The rest of the population falls in two tails: one with an apocentre at 10\,au, and another, less well-defined, with a pericentre at $\sim15$\,au. These are respectively the initial semi-major axis of the inner planet, and the average initial semi-major axis of all planets. The region in between is underpopulated, since planets here in general are still experiencing encounters and hence systems are unstable \citep{Chatterjee+08}.

\begin{figure}
  \includegraphics[width=0.5\textwidth]{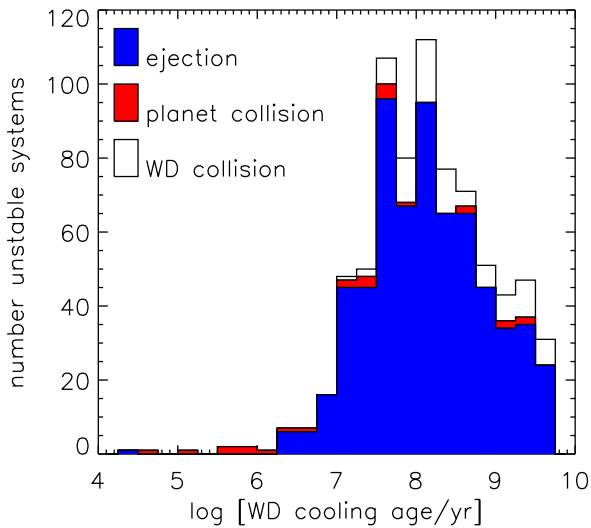}
  \caption{Cooling ages at which WD instabilities occur in systems of three $1\mathrm{\,M_J}$ planets, and their instability outcome.}
  \label{fig:tcool}
\end{figure}

Instabilities during the WD phase tend to be delayed. Figure~\ref{fig:tcool} show a histogram of the cooling ages of the WD at which planets are lost, summed across all stellar masses but broken down by the result of the instability. The ejections, forming the bulk of the instabilities, mostly occur after 1\,Myr and continue until the end of the integrations. It appears that ending the integrations at 5\,Gyr has truncated the tail of the distribution and instabilities would still occur, albeit less frequently, at still later ages. The planet--planet collisions begin much earlier, and their rate per logarithmic time unit is roughly constant. Collisions with the WD begin much later, at 10\,Myr, and there is a hint of bimodality in the distribution of the logarithm of their ages. There is also a difference in the times at which the systems that did or did not experience pre-WD planet losses experience instability, with a median of $10^{7.9}$ years for the first WD instability in hitherto stable systems, versus $10^{8.4}$ years for the first WD instability in previously unstable systems. A KS test shows that the two distributions are significantly different ($p=5\times10^{-8}$). Hence, systems which had previously experienced instability see their WD instabilities occurring a little later than the stable systems.

In Figure~\ref{fig:a,e,q} we show the final orbital elements of all systems at 5\,Gyr, as plots of eccentricity $e$ and pericentre $q$ against semi-major axis $a$. Planets in stable systems, marked in red, have semi-major axes in the range 40--150\,au, as expected from the adiabatic expansion of their initial orbits, and low eccentricities which have been somewhat excited by planet--planet interactions. Unsurprisingly, planets in unstable systems have wildly different orbital elements, with semi-major axes up to several thousands of au and eccentricities up to 1. The distribution of elements in unstable systems is similar to that in Figure~\ref{fig:a-e-ms}, but the structure is somewhat washed out by the different amounts of orbit expansion on the AGB. A particularly dense concentration can be found with semi-major axes of 20--30\,au, smaller than the minimum adiabatically-expanded initial value. Few semi-major axes are lower than these values. This is in accordance with the finding of \cite{Chatterjee+08} that scattering does not reduce semi-major axes below about 40\% of their initial value. Nonetheless, some of these planets can approach closer to the star, since they have very high eccentricity and hence very small pericentres. The survivors' pericentres at 5\,Gyr are shown in the lower panel of Figure~\ref{fig:a,e,q}, where we see that some planets come within 10\,au of their host star, a region expected to be cleared during the AGB phase \citep{MV12}. The structure of the $a-e$ distribution will be somewhat sensitive to initial conditions, since the semi-major axis of the inner planet sets the scale for the major concentrations of points, as discussed for Figure~\ref{fig:a-e-ms} above. However, the existence of a hard inner cut-off in $a$ will be robust, since if planets are started initially too close to the star, they will be swallowed by the stellar envelope. We note that three planets classed as being in ``stable'' systems clearly belong to the unstable population from their orbital elements, with $a\approx30, 350$ and $400$\,au. These planets all belong to one system, which underwent a scattering event at $\sim4.975$\,Gyr, just before the end of the integration. As we have no way of telling what the outcome of this instability will be, we choose to keep the system classified as ``stable''.

\begin{table}
\begin{center}
\begin{tabular}{ccccccc}
                             &                            & \multicolumn{4}{c}{number of planets}\\
$M_\mathrm{pl}/\mathrm{M_J}$ & $M_\star/\mathrm{M}_\odot$ & 0 & 1  & 2   & 3 & number systems\\
\hline
1                            & 3                          & 2 & 62 & 71  & 113 & 248\\
                             & 4                          & 0 & 68 & 80  & 100 & 248\\
                             & 5                          & 3 & 78 & 81  & 86 & 248\\
                             & 6                          & 2 & 72 & 109 & 65 & 248\\
                             & 7                          & 1 & 81 & 83  & 83 & 248\\
                             & 8                          & 1 & 78 & 88  & 81 & 248\\
10                           & 3                          & 6 & 50 & 71  & 121 & 248
\end{tabular}
\caption{Number of systems with specified surviving number of planets at 5\,Gyr, by stellar and planetary mass. There are 248 systems for each row.}
\label{tab:nsurv}
\end{center}
\end{table}

We have seen that systems may lose several or even all three planets during the course of stellar evolution. Table~\ref{tab:nsurv} shows the number of surviving planets at 5\,Gyr in all our simulation sets. Nearly all WDs that initially had systems of giant planets on wide orbits will keep at least one. In general, the more massive a star, the more systems will be destabilised and the fewer planets will be found around WDs. We discuss the observational consequences of this in Section~\ref{sec:discussion}.

The planetary systems of WDs are more unstable than those around MS stars for one simple reason: the increase in the planet:star mass ratio, causing a decrease in the planetary separation when measured in Hill radii. Several terms in Equation~\ref{eq:delta-postms} have a smaller effect. The decrease in stellar mass directly causes an increase in the orbital time-scale, and hence any process that takes a specified number of orbits would take a larger amount of real time. The increase in the planets' semi-major axes causes a further increase in the orbital period, compounding this effect. However, since time and semi-major axis only affect the critical separation logarithmically, the power-law dependence of the Hill radius on the stellar mass is the dominant effect.

As we showed in Paper~I, systems of two planets are also susceptible to instability following mass loss. A notable difference when considering three-planet systems is the potential to undergo two rounds of instability: a first round on the MS which causes the loss of one planet while the survivors are left in a stable state, and then a second round as a two-planet system as a WD. Indeed, systems that had been already destabilised on the MS were just as likely in our simulations to then lose a second planet as a WD than those systems that survived from the MS intact. While in both cases ejections dominate as an outcome of WD instability, in our three-planet simulations we see far more collisions with the star, and far fewer planet--planet collisions, than in Paper~I. Rather than being a fundamental change in the underlying dynamics, this likely reflects our attempts to improve the realism of our simulations: by assigning the planets a small initial inclination we greatly reduced the chances of a planet--planet collision compared to the coplanar simulations of Paper~I, while increasing the radius for stellar collisions from the physical WD radius to the Roche limit has greatly increased the number of planets classed as ``colliding with the star''. The precise fate of such bodies will be discussed below.

\begin{figure}
  \includegraphics[width=0.5\textwidth]{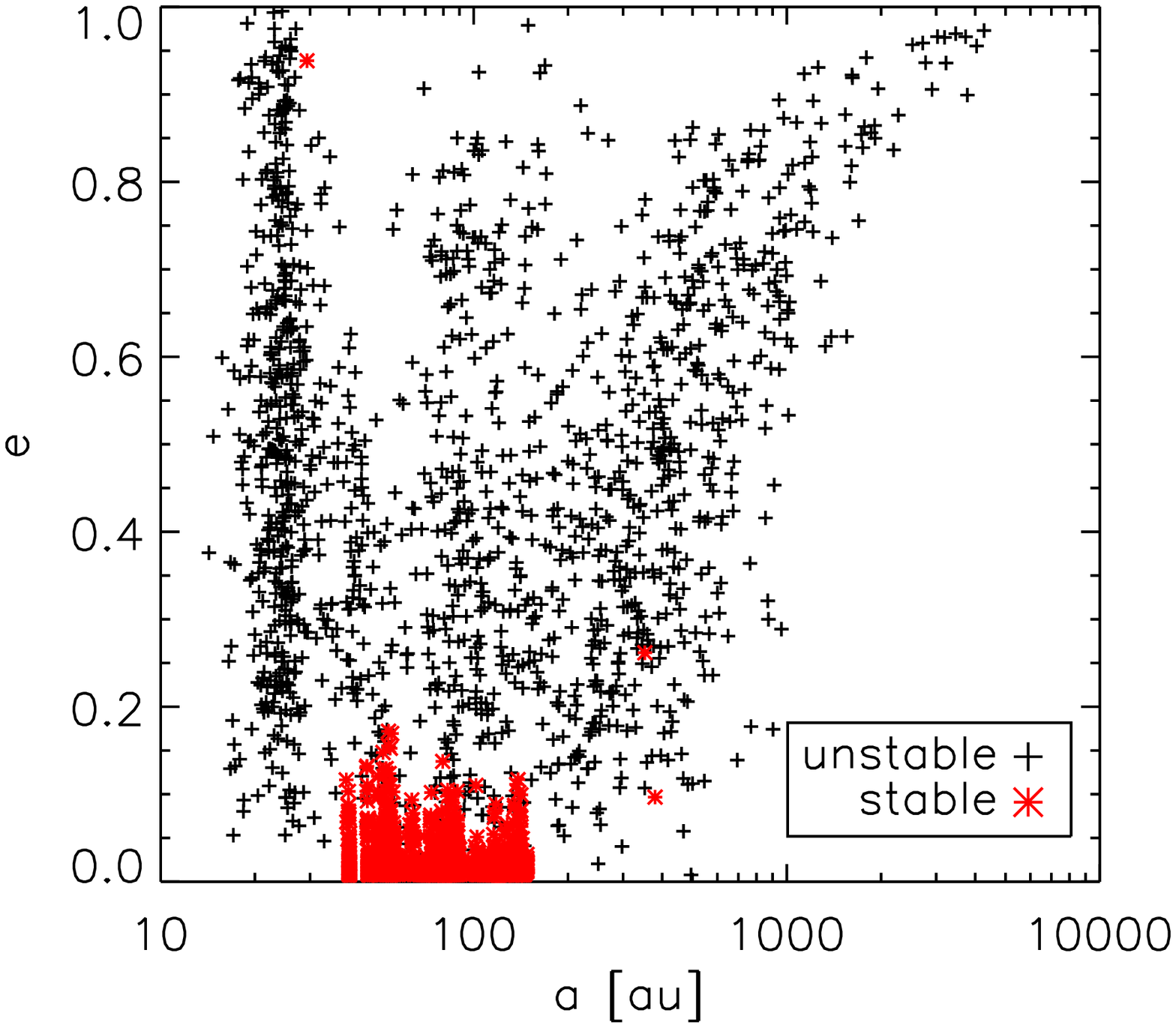}
  \includegraphics[width=0.5\textwidth]{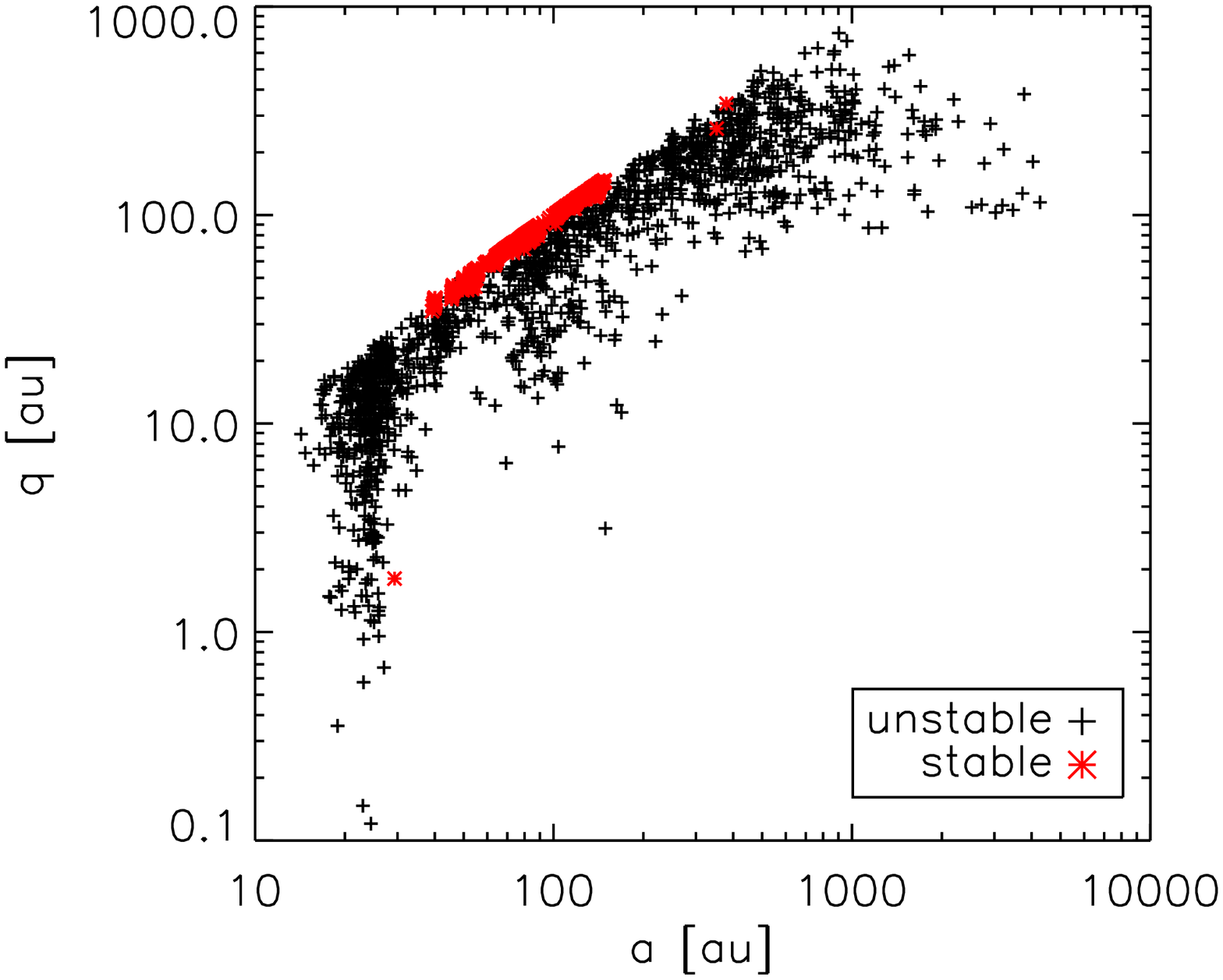}
  \caption{Final  (at 5\,Gyr of total evolution, \emph{i.e.,} more than 4\,Gyr into the WD stage) eccentricities (top) and pericentres (bottom) as a function of final $a$. Planets in systems that have remained stable for 5\,Gyr are marked in red; those in systems that have experienced an instability, in black.}
  \label{fig:a,e,q}
\end{figure}

\subsection{High planet mass case, $M_\mathrm{pl}=10\mathrm{\,M_J}$}

\label{sec:10mj}

\begin{figure}
  \includegraphics[width=.5\textwidth]{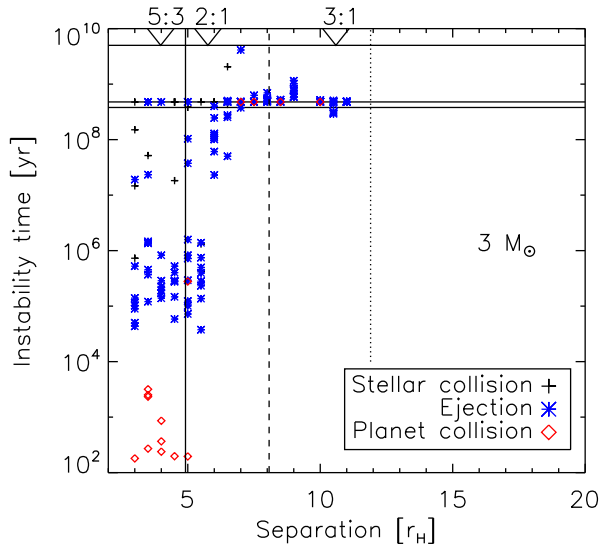}
  \caption{As Figure~\ref{fig:times-by-type} for the case of three $10\mathrm{\,M_J}$ planets.}
  \label{fig:tstab-10mj}
\end{figure}

We now discuss the case of three $10\mathrm{\,M_J}$ planets. The times and types of instability are shown as a function of initial separation in Figure~\ref{fig:tstab-10mj}, while the planet losses by evolutionary stage are listed in Table~\ref{tab:outcomes-10mj}. Ejections dominate the outcomes during the MS (62 of 81 planets lost, $76.5^{+4.0}_{-5.3}\%$). On the MS, systems are unstable out to $\sim6.5r_\mathrm{H}$, while the semi-analytical estimate from Equation~\ref{tinst} predicts that the stability boundary should be at $8r_\mathrm{H}$. There is also a small region of instability associated with the 3:1 MMR at $\sim10.5r_\mathrm{H}$. Hence, the predictions of the semi-analytical estimates are not so good as in the lower planet mass case, possibly because the formulae have not been calibrated at such high mass ratios. In contrast, the Hill criterion applied pairwise predicts that planet--planet collisions can occur out to $5r_\mathrm{H}$, which is indeed borne out by the integrations. 

As in the case of the $1\mathrm{\,M_J}$ systems, planet losses between the end of the MS and the end of the AGB are primarily collisions with the expanded stellar envelope (19 out of 26, $73.1^{+6.8}_{-10.2}\%$). Again, losses primarily occur in systems that had previously lost a planet, with 3 planets lost being the systems' first ($11.5^{+9.2}_{-3.6}\%$), 17 being their second ($65.4^{+7.8}_{-10.1}\%$), and 6 their third ($23.1^{+10.0}_{-6.2}\%$).

\begin{figure}
  \includegraphics[width=.5\textwidth]{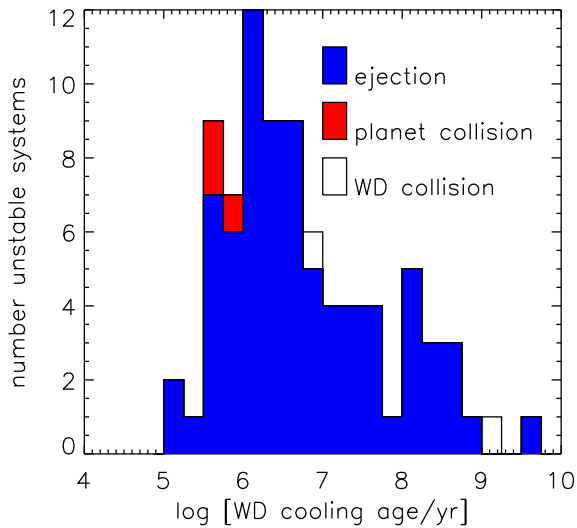}
  \caption{Histogram of cooling ages at which $10\mathrm{\,M_J}$ planets are lost around WDs.}
  \label{fig:tcool-10mj}
\end{figure}

When compared with the $1\mathrm{\,M_J}$ case, destabilisation of systems round WDs remains common. Numerically, we find that systems can be unstable out to a separation of $11r_\mathrm{H}$, close to the $12r_\mathrm{H}$ predicted from Equation~\ref{eq:delta-postms}. However, this may be due to the enhanced destabilising effects of the 3:1 MMR.

The nature of planet loss around WD stars in the high planet mass case is not the same as for lower planet masses. Of 182 three-planet systems entering the WD stage, 61 experienced instability ($33.5^{+3.7}_{-3.3}\%$). However, only 4 of the 31 systems of two planets at the start of the WD stage lost a further planet ($12.9^{+8.4}_{-3.9}\%$). This contrasts with the overall results from the lower planet mass case, where already destabilised systems were as likely to lose an additional planet as intact ones; although, when considering only $3\mathrm{\,M}_\odot$ primaries with $1\mathrm{\,M_J}$ planets, the difference is less noticeable (here, $31.5^{+3.8}_{-3.4}\%$ of intact systems lost a planet, compared to $20.0^{+7.7}_{-4.9}\%$ of previously destabilised ones). The times of instability are also shifted to much earlier ages (Figure~\ref{fig:tcool-10mj}), with the median age of loss at $10^{6.5}$ years, compared to $10^{8.3}$ years, the distributions differing significantly (KS test $p$-value of $3\times10^{-14}$). As in the lower mass case, planet--planet collisions tend to occur earlier and planet--WD collisions later, although the numbers of such collisions are small. Indeed, ejection is the most common outcome by far, with 77 of the 82 planets lost ($93.9^{+1.7}_{-3.8}\%$) being ejected.

\begin{table}
\begin{center}
\begin{tabular}{llcccc}
                      &       & MS  & early-PMS & WD  & Total\\\hline
$3\mathrm{\,M}_\odot$ & SC    & 8   & 19        & 2   & 29\\
                      & EJ    & 62  & 6         & 77  & 145\\
                      & PC    & 11  & 1         & 3   & 15\\
                      & Total & 81  & 26        & 82  & 189
\end{tabular}
\caption{Number of planets lost in the three-$10\mathrm{\,M_J}$ runs, broken down by type of loss (``SC'' = stellar collision, ``EJ'' = ejection, ``PC'' = planet--planet collision) and stellar evolutionary phase (``MS'' = Main Sequence, ``early-PMS'' = subgiant through to end AGB, ``WD'' = White Dwarf).}
\label{tab:outcomes-10mj}
\end{center}
\end{table}

\section{Discussion}

\label{sec:discussion}

\subsection{Generality and reality of our simulations}

\label{sec:generality}

As the expensive nature of N-body integrations only permits a partial exploration of parameter space, we need to address two questions. First, how general are our results? And second, how representative are our systems of those occurring in reality?

Can we generalise the results of our simulations to other values of the planet and stellar mass, and initial semi-major axes, using the semi-analytical scalings presented in Section~2? We restricted our integrations to the study of $1\mathrm{\,M_J}$ planets around $3-8\mathrm{\,M}_\odot$ progenitor stars and $10\mathrm{\,M_J}$ planets around $3\mathrm{\,M}_\odot$ progenitor stars, with the inner planet initially at 10\,au. For the lower mass planets, for all the stellar masses which we explored in our N-body integrations, the estimated stability limits from Equations~\ref{tinst} and~\ref{eq:delta-postms}, which were based on the work of \citep{FaberQuillen07} who studied planet:star mass ratios in the range $10^{-7}-10^{-3}$, gave a good description of where systems can be unstable for MS and WD primaries, as can be seen in Figure~\ref{fig:times-by-type}. When counting the star's mass loss, this covers more than a factor of 10 in planet:star mass ratio and 5 in the inner planet's semi-major axis. While these ranges are small compared to the full possible range of planetary masses and semi-major axes, they do inspire confidence that the estimates in Section~2 can be extrapolated to lower planet and stellar masses and to different semi-major axes. We do note that the $10\mathrm{\,M_J}$ simulations are rather more stable than the analytical estimates predict. However, below we only consider extrapolation to lower planet masses.

\begin{figure}
\includegraphics[width=.5\textwidth]{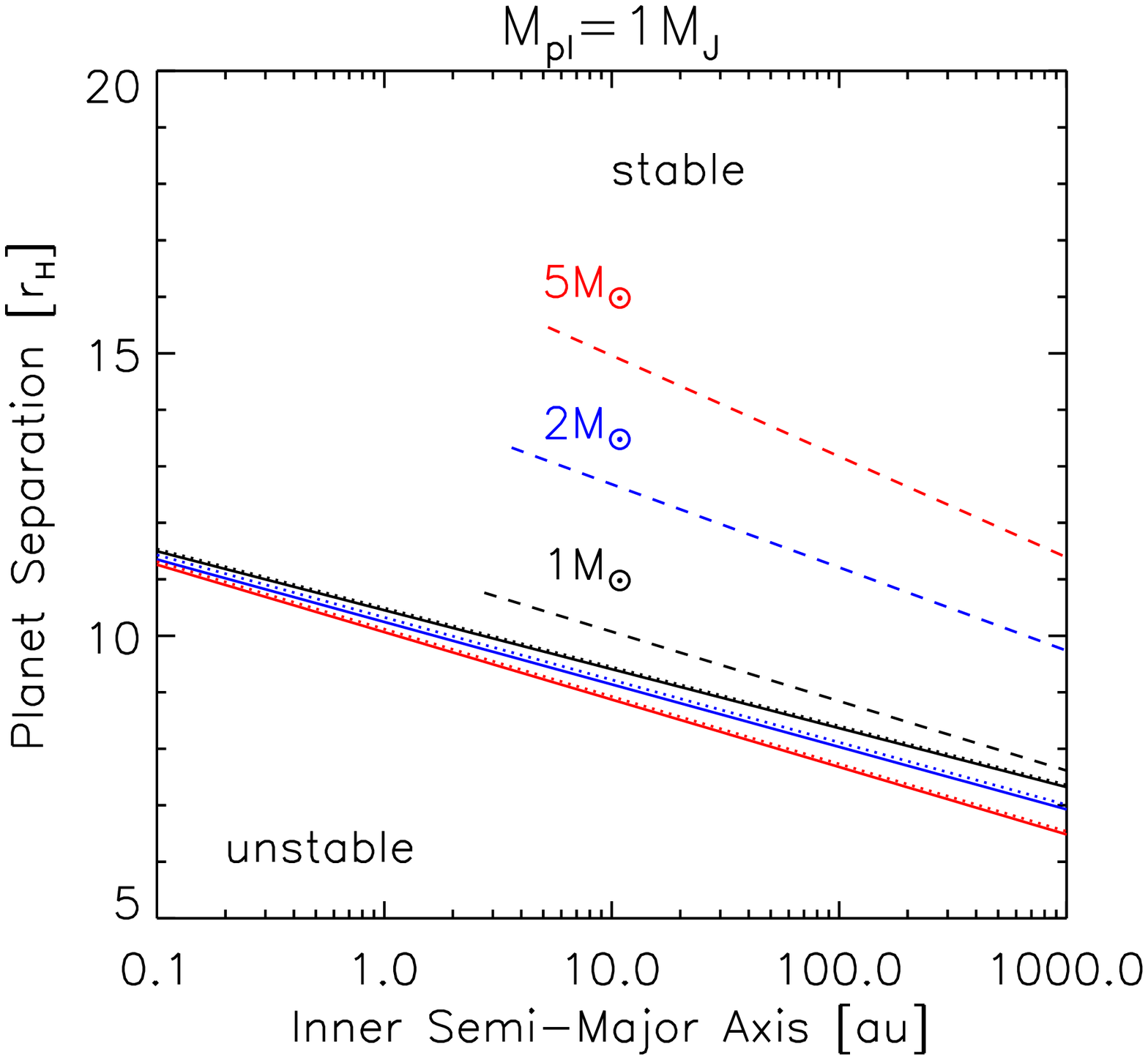}
\includegraphics[width=.5\textwidth]{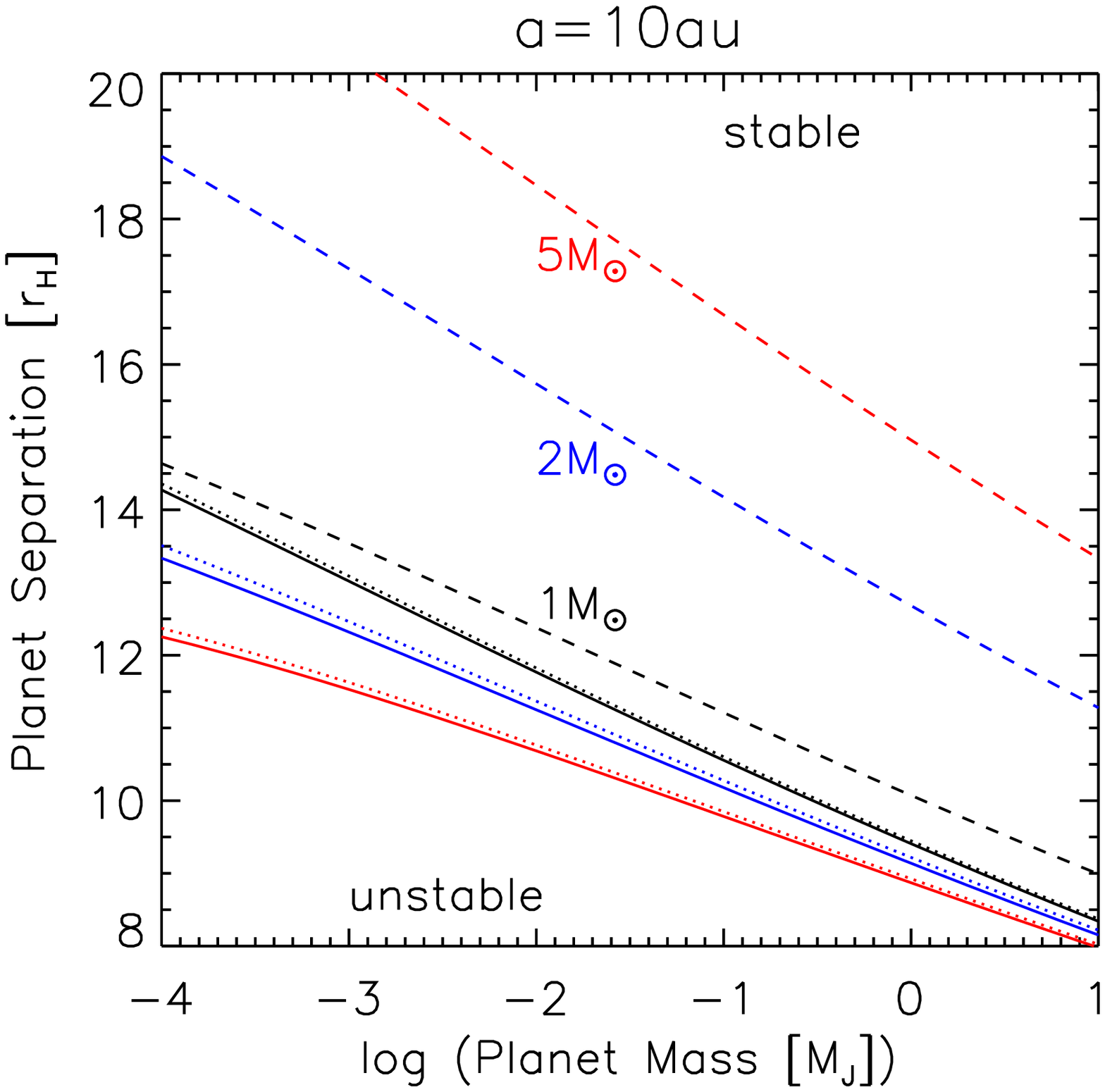} 
\caption{
Dependence of the estimated stability limits on inner planets' semi-major axes 
({\it upper panel}) and masses ({\it lower panel}) for different
progenitor stellar masses, shown in different colours. Systems below the solid lines are unstable on the MS; between the solid and the dotted line they are unstable between the end of the MS and the onset of AGB mass loss; between the dotted and dashed lines they are unstable around WDs; and above the dashed lines they are stable for the age of the Universe. In the upper panel, the
the planetary masses are each $1 M_J$. In the lower panel,
the innermost semi-major axis is $10$ au. Upper and lower solid lines show the stability limits at the end of the MS and at a total age of $13.7$\,Gyr. Dotted lines show the stability limits just before mass loss on the AGB. In the upper panel, the WD stability limits are truncated at the \emph{moriturus ultimus} radius of Mustill \& Villaver (2012), within which the inner planet will be engulfed by the AGB star.
}
\label{fig:stable-est}
\end{figure}

Assuming that Equations~\ref{tinst} and~\ref{eq:delta-postms} accurately reflect actual stability properties,, we can predict the range of stable and unstable planetary separations as a function of the inner planet's semi-major axis and the planetary and stellar mass. These ranges are shown in Figure~\ref{fig:stable-est}, for planets around 1, 2 and $5\mathrm{\,M}_\odot$ stars. Less massive planets, and those at smaller radii, require a larger separation in Hill radii in order to be stable over any given time period. However, considering the ranges of parameters considered, the critical separations are remarkably insensitive to the planets' masses and semi-major axes. In contrast, reducing the stellar mass to smaller values has a large effect: the increase in the number of systems that are unstable during the WD phase will be significantly smaller at lower stellar masses, as a consequence of the smaller fraction of mass lost from the star. There will also be an increase in the number of unstable systems on the MS at lower stellar masses, a trend which is seen in our integrations (see Table~\ref{tab:outcomes}). We note that for Solar mass stars significant mass loss can occur on the RGB as well as the AGB, and this may induce orbital instability during the core Helium burning phase.

We have considered an idealised case where three planets of the same mass form on orbits separated by a constant semi-major axis ratio. In general, systems will not form in such a tidy manner. However, \cite{Chambers+96} showed that including a modest spread of planet masses or semi-major axis ratios does not significantly change the time-scale for instability in a system, although some features such as the effects of MMRs become less pronounced. We do note, however, that the outcomes of instability, including the effects of a second round of instability following mass loss, will be affected. For example, reducing the mass of one planet in a system will increase the likelihood of its being ejected, as its binding energy is lower.

More generally, we acknowledge that the statistics on instability occurrence and outcomes given in Section~\ref{sec:sims} may not exactly relate to what will happen in real systems. Even if all systems were indeed of equal-mass, equally-spaced planets as we have assumed, the relative frequencies of instabilities around stars at different ages would still depend on the initial separation of the planets. This must be borne in mind when interpreting our statements such as ``The number of planets lost on the WD exceeds the number lost on the MS in all but the $3\mathrm{\,M}_\odot$ case.'' If planets formed preferentially on very packed orbits, the frequency of MS instability would rise relative to that seen in our simulations.

We must therefore consider more carefully the relationship of our simulations to real planetary systems. Unfortunately, the statistics on systems similar to the ones we study are poor. First, most host stars are smaller than those we consider, with few having masses above $3\mathrm{\,M}_\odot$. Second, we are considering relatively large semi-major axes and most detection techniques are biased towards small semi-major axes, with only direct imaging being able to probe the separations we consider. Several planetary systems have now been imaged around stars more massive than the Sun, at wide separations. \cite{Vigan+12}, in a survey of 42 MS AF stars, find two with super-Jupiter planets (above $3\mathrm{\,M_J}$), implying a fraction of $5.9-18.8$\% having such planets between 5 and 320\,au after correcting for sensitivity. This limit is consistent with the determination by \cite{Nielsen+13} that $\lesssim20\%$ of $1.5-2.5\mathrm{\,M}_\odot$ stars host $4\mathrm{\,M_J}$ planets, and $\lesssim10\%$ host $10\mathrm{\,M_J}$ planets, at separations of tens to hundreds of au. One of the \cite{Vigan+12} systems is the famous multiple-planet system HR~8799 \citep{Marois+08,Marois+10}, which is the closest match to our simulated systems amongst those known. This suggests that multi-planet systems such as we have considered do indeed exist, although they might be uncommon.

Further estimates of the occurrence of the planetary systems herein considered must rely on extrapolations from closer-in planets detectable by radial velocity measurements or on models of planet formation. From an RV survey of 31 subgiants, \cite{Bowler+10} find the occurrence rate of giant planets orbiting within 3\,au of evolved intermediate-mass ($1.5-2\mathrm{\,M}_\odot$) stars to be $26^{+9}_{-8}\%$. \cite*{Maldonado+13} find that planet-hosting giant stars of $\gtrsim1.5\mathrm{\,M}_\odot$ show metal enrichment similar to MS planet hosts, suggesting similar formation mechanisms around stars of different masses. From a theoretical perspective, \cite{KennedyKenyon08} argue that giant planet formation through core accretion should be at its most efficient at stellar masses around $3\mathrm{\,M}_\odot$: protoplanetary disc mass may increase with stellar mass, encouraging core formation, but the discs of more massive stars are short lived, decreasing the time available to form cores. However, planet formation through gravitational instability will not be subjected to the same time-scale restrictions and hence may not be disfavoured around higher-mass stars.

\subsection{WD planets on wide orbits}

We now turn to discuss the implications of our study for searches for planets around WDs. Unless planets can form from material ejected from the AGB star, planets around WDs must have survived the full evolution of the star since they formed in the protoplanetary disc. Planets initially within a few au of the star will be engulfed and probably destroyed during the star's AGB evolution \citep{VL07,VL09,MV12}, although bodies in the Brown Dwarf mass range may survive common envelope evolution to end up on very tight orbits around the WD \citep{Maxted+06,Parsons+12,NS13}. \cite{MV12} showed that giant planets that survive engulfment by the star must be found at distances beyond 2\,au, for a $1\mathrm{\,M}_\odot$ progenitor, rising to 10\,au for a $5\mathrm{\,M}_\odot$ progenitor. How does the incorporation of multi-planet interactions change the expected distribution of WD planets?

First, a very few systems lost all planets during the course of their evolution. While all single-planet systems where the planet is in an orbit sufficiently wide to avoid engulfment in the envelope will survive to the WD, 15 of our 1736 multi-planet systems simulated ended up devoid of planets. However, neglecting this small rate of total planet loss, it will be the case that every intermediate-mass MS star that hosts at least one wide-orbit ($\gtrsim10$\,au) giant planet should give rise to a WD which also hosts at least one wide-orbit giant planet. Thus, the fraction of wide-orbit planets around MS stars and WDs should be identical.

While there have been several surveys for planets around single WDs, there are currently no firm detections and even the upper limits on the occurrence rate are as yet fairly weak. Direct detection of planets' thermal emission is aided by the low luminosity of WDs compared to MS stars, and by the shift of the stellar spectrum towards shorter wavelengths \citep*{Burleigh+02}. However, the age of systems, typically $>1$\,Gyr, means that planets and BDs have cooled significantly, making only the most massive planets detectable (although planets can experience significant heating during the planetary nebula phase which enhances the luminosity of planets orbiting young WDs, \citealt{VL07}). In searches for unresolved companions, \cite{Farihi+08} and \cite{Kilic+09} found no evidence for IR excesses consistent with unresolved $10\mathrm{M_J}$ companions in a sample of 40 WDs, implying a $2\sigma$ upper limit of $<7$\%. Unfortunately the limits for Jovian-mass objects are very weak. Both of these surveys focussed on WDs with progenitor masses $\gtrsim3\mathrm{\,M}_\odot$. \cite{Debes+11} reported a low fraction (1--5\%) of candidate BD companions to WDs, but background confusion prevented secure identification of these sources. While these studies focussed on unresolved companions, \cite{Hogan+09} sought common-proper motion companions to 23 WDs, putting a limit of $\lesssim9$\% on the fraction of WDs hosting $10\mathrm{\,M_J}$ planets at separations of 60--200\,au.

\cite{Kilic+09} compared their upper limit on planets orbiting WDs to the detection rate of closer-in RV-detected planets round intermediate-mass ($1.3-1.9\mathrm{\,M}_\odot$) SG stars ($8.9\%$, reported in \citealt{Johnson+07}), finding that the non-detections around WDs are consistent with the occurrence rate around MS stars, but that they are suggestive of a lower frequency. With the more recent results of \cite{Bowler+10} described above, this discrepancy becomes larger. \cite{Kilic+09} suggest three reasons for this difference: first, stars of more than $3\mathrm{\,M}_\odot$ may be inefficient at forming planets \citep{KennedyKenyon08}; second, planets may be lost due to the direct effects of stellar evolution, particularly on the AGB \citep[e.g.,][]{VL07}; third, planets may be lost through instabilities induced through stellar mass loss \citep{DS02}. They discount the second possibility's being important for their sample as planets on wide orbits are safe from the effects of photoevaporation and tides. As a result of our simulations, we have shown that the third reason should not cause a large difference between occurrence rates of planets around MS and WD stars: In only 15 of our integrations, just 1\%, did the system lose all of its planets during stellar evolution, despite instability being so common, and in non of these cases was the final planet lost as a result of instability during the WD stage. Hence, we should find wide-orbit planets at similar rates around both MS and WD stars. 

We caution that the IR surveys relying on IR excess are biased against wider companions, and comparisons of planet occurrence rates will have to take into account the often large expansion of planetary orbits. In Table~\ref{tab:outer} we show the number of systems in our simulations whose only surviving planets' semi-major axes were outside a specified radius. In the most extreme case, $47.8\pm1.3$\% of the $1\mathrm{\,M_J}$ systems had no planets surviving within 50\,au, which may not show up as unresolved IR excesses and would require multi-epoch imaging. The exact numbers given in this table will be rather sensitive to our initial planet masses and semi-major axes, but it will remain the case that in some systems the surviving planet(s) end up on fairly wide orbits around the WD. While planets from single progenitor systems are not expected to be found within a few 10s of au of massive WDs \citep{VL07,MV12}, scattering in multiple systems can clear planets from much wider regions.

\begin{table}
\begin{center}
\begin{tabular}{p{.6cm}p{.7cm}p{.7cm}p{.5cm}p{.5cm}p{.5cm}p{.5cm}p{.5cm}}
         Planet      & Initial & Surviving &&&&&\\
         mass        & systems & systems   & 50\,au & 100\,au & 200\,au & 500\,au & 1000\,au\\\hline
  $1\mathrm{\,M_J}$  & 1488    & 1479      & 707    & 254     & 143     & 76      & 27\\
  $10\mathrm{\,M_J}$ & 248     & 242       & 28     & 24      & 21      & 7       & 2
\end{tabular}
\caption{Number of systems with one or more planets surviving to WD stage, where the only surviving planets have semi-major axes greater than that specified.}
\label{tab:outer}
\end{center}
\end{table}

\subsection{WD planets on close orbits}

There have also been searches for planets close to the WD. \cite{Faedi+11} searched for transits of 194 WDs in the WASP survey, finding an upper limit of 10\% for the fraction of WDs hosting very short-period ($<0.2$\,day) giant planets, and progressively weaker limits at larger orbital distances. It is unlikely that planets below $\sim10\mathrm{\,M_J}$ can survive engulfment in the stellar envelope during the AGB \citep{VL07,NS13}, although the existence of planets around a horizontal branch star \citep{Charpinet+11} suggests that in some cases planets can survive, or reform following tidal disruption \citep{BS12}. Hence, close-in planets around WDs must have been brought to the WD vicinity after the WD formed, likely through dynamical processes followed by tidal circularisation of their orbit. Our simulations see a small fraction of systems having planets which pass within the WD's Roche limit: 85 planets, out of 755 lost during the WD evolution, in our fiducial $1\mathrm{\,M_J}$ case, in 82 separate systems ($5.5^{+0.6}_{-0.5}\%$ of all systems). We can make a simple physical argument as to why the delivery of planets close to the WD from these large distances is unlikely. Consider a simple two-planet system where both planets are at approximately the same semi-major axis $a^\mathrm{(i)}$ and have the same mass. If one planet acquires just enough angular momentum to become unbound, its angular momentum increases by a factor of $\sqrt{2}$. The survivor's angular momentum after ejection is therefore $L^\mathrm{(f)}=(2-\sqrt{2})L^\mathrm{(i)}$, so it has lost angular momentum during the encounters. However, its apocentre $Q^\mathrm{(f)}$ will be at approximately the original semi-major axis $a^\mathrm{(i)}$, and so the final eccentricity would be around $0.66$, meaning a pericentre of 8\,au for a planet with an apocentre at 40\,au. In numerical simulations, \cite{FR08} found a maximum eccentricity of $\approx0.8$ for the survivor of a two-planet system. More complicated exchanges of angular momentum in a three-planet system can of course take place, but it remains challenging to drive a planet's eccentricity to the very high values ($\gtrsim0.99$) required to be able to tidally circularise it.

The actual fate of these planets may vary. In the next subsection we discuss the effects of their tidal disruption. Here, however, we ask what happens if, before the planets are driven inside the Roche limit, they first pass the star at a slightly greater distance and their orbits become tidally circularised. This would result in their being dynamically decoupled from the source of perturbations that would have eventually driven them into the star, while eventually shrinking the orbit into a circular one at just beyond $2R_\mathrm{Roche}$. This mechanism is proposed to explain the distribution of hot Jupiters around MS stars \citep*{FabryckyTremaine07,Nagasawa+08}, and the dynamics around WDs will be the same, since the dominant tidal forces occur in the planet and do not depend on the evolutionary state or structure of the star.

As shown in Figure ~\ref{fig:a,e,q}, surviving planets at the end of our integrations in general do not have pericentres close to the Roche limit. Because the pericentres evolve with time, some may come closer than recorded at the end of the simulation. Over the whole WD stage, we found 76 planets with a pericentre recorded within 0.1\,au of the WD, of which 48 later collided with the star. Although our coarse time sampling (every $1$\,Myr) means that we may have missed some planets that have close approaches and may be tidally circularised, these figures suggest that the bulk of potentially close-in planets come from those that would suffer a stellar collision in the absence of tides.  Hence, we assume the maximum fraction of systems forming ``hot Jupiters'' around WDs would be given by the fraction that eventually enter the Roche limit, $\sim5.5$\%. Since $\lesssim20$\% of MS systems host the systems of massive planets that we consider in this paper, this would set an upper limit of $\lesssim1$\% of WDs hosting close-in Jupiters, which then would require a favourable geometric alignment to be observed as transiting systems. This is consistent with the upper limit of 10\% of WDs hosting close-in Jupiters from \cite{Faedi+11}, and we can predict that we can only expect detections of such planets with greatly increased sample sizes of several thousands of WDs.

The number of planets scattered in does of course depend on the system architecture. In our $10\mathrm{\,M_J}$ simulations, only $0.8^{+1.0}_{-0.3}$\% of systems saw a planet driven to the Roche limit, meaning the number of transiting super-Jupiters will be even smaller. However, reducing the planet mass considerably will not result in many bodies being scattered in: in our two-planet simulations in Paper~I, for the case of two Earth-mass planets with the inner at 10\,au, we saw no collisions with the star over the whole stellar lifetime, including during the giant stages ($<0.6\%$). Although there was no strong trend in the number of collisions with the star as a function of stellar mass, it is the case that smaller stellar masses allow survival closer to the star during the AGB, and scattering from tighter final orbital radii may increase the number of close-in planets slightly. On the other hand, as discussed above, the extra parameter space destabilised is likely to be smaller. Hence, lower-mass WDs with progenitors in the range $1-3\mathrm{\,M}_\odot$ may offer a higher probability of hosting transiting planets, but this will still likely be small.

\subsection{WD Pollution}

If a planet is not tidally circularised before reaching the Roche limit, it will suffer partial or complete disruption \citep{Zuckerman+07,Klein+10,Debes+12,Gaensicke+12}. The dynamics of planets passing within or close to the Roche limit are complex and depend on the planet's internal structure and the distance of approach to the star. If total disruption does not occur, mass is stripped from the outer layers of the planet, the remainder changing its orbit to become either more or less bound to the star, or even in some cases unbound \citep{Faber+05,Guillochon+11,Liu+13}. A delivery of all of a planet's mass to pollute a WD would represent only an extreme case.

Direct pollution of a WD in this way is however rare: up to $5.5$\% of massive WDs with multi-planet systems may be polluted directly by a gas giant. The accretion of a $1\mathrm{\,M_J}$ planet onto a WD would lead to a peak luminosity of order $10^4\mathrm{\,L}_\odot$, using the scalings in \cite{BS13}. However such events would occur infrequently. Assuming a Galactic population of $\sim10^{10}$ WDs \citep{Napiwotzki09}, 10\% of which have multi-giant planet systems, and considering that 5\% of these will send a planet into the WD over a 5\,Gyr time period, suggests that the rate of WDs accreting Jovian planets should be about one per century in the Galaxy.

Observed signatures of metal pollution in WDs, which occurs in around 25\% of DA WDs \citep{Zuckerman+03}, are generally consistent with the accretion of rocky asteroids. \cite{DS02} proposed that the destabilisation of giant planet systems could trigger the destabilisation of planetesimals in the system and their subsequent accretion onto the WD. Our integrations confirm that instability is a common outcome for closely-packed multi-planet systems around WDs. The time at which the instabilities occur in these integrations is at cooling ages of several 100\,Myr, which was also the case in the two-planet simulations we studied in Paper~I. As discussed in that paper, this is similar to the ages at which metal pollution is observed. However, the low fraction of MS stars and WDs that host these multi-giant planet systems shows that such instabilities cannot occur around a large fraction of WDs. Moreover, around a quarter of planetary systems that experience instability around WDs have previously experienced instability, which would have already depleted their planetesimal reserves, although such systems may still see low levels of pollution from residual planetesimals. If we take an observational upper limit of $20$\% of stars hosting systems of multiple giant planets on wide orbits, as discussed in Section~\ref{sec:generality}, and consider that in our integrations 43\% of WD systems are unstable, we find that giant planet instability alone would predict pollution of $\lesssim9$\% of WDs. This falls somewhat short of the observed pollution occurrence fraction; furthermore, studies of planetesimal scattering in the wake of planetary instability around MS stars show that a high planetesimal flux to the regions near the star lasts only a short time \citep{Booth+09,Bonsor+13}. Pollution can be achieved with a single planet, although the required planetesimal reservoirs are often orders of magnitude larger than the Solar System's asteroid belt \citep{Debes+12}. We note that a collision between terrestrial planets may create a fresh reservoir of debris during the WD phase that could then be perturbed by other planets (B.~G\"ansicke, priv.~comm.). While planet--planet collisions were not a common event during the WD phase in the simulations described in this paper, we did not study terrestrial-mass planets. In Paper~I we found that planet--planet collisions were more common in systems of two Earth-mass planets than in systems of two giant planets. Further studies of systems including low-mass planets would be needed to quantify this process and its sensitivity to the system set-up. It is likely that a variety of mechanisms plays a role, commensurate with the diversity of planetary system architectures.

\subsection{Free-floating planets}

A purportedly vast population of free-floating giant planets is travelling between and outnumbers the stars of the Milky Way, as revealed by microlensing observations \citep{sumetal2011}. If sufficiently young and nearby, free-floating planets may be imaged. Although the mass of these free-floaters is uncertain due to a degeneracy in their age, luminosity and mass, \cite{deletal2012} have recently provided a near-confirmation of the mass for one of these objects, helping to confirm the existence of free-floating giant planets.

The source of such a large free-floating planet population is unlikely to be dynamical instabilities arising from gravitational scattering alone because otherwise the average number of giant planets formed per planetary system would have to be unrealistically high \citep{verray2012}.  However, that study considers scattering only on the MS, and claims that scattering during the WD phase contributes to the total free-floating giant planet population on the order of 1\% based solely on the currently observed space density of WDs.

Our work can finally help assess the relative contributions to the free-floating population from MS scattering versus WD scattering.  Although the fraction is dependent on our initial conditions, our initial condition choices encompass the entire phase space of separations from near-guaranteed instability to near-guaranteed stability.  Further, our assumption of initially circular orbits essentially controls the time-scale for instability to occur, and hence might have a weak dependence on the resulting dynamics for ensembles of systems \citep[e.g.][]{JuricTremaine08}.

We find that, summed over all simulations reported in Table 1, an ejection is 3.9 times more likely to occur while the parent star is a WD than when the star is a MS star.  This difference is most pronounced for the highest mass stars (with a factor of $8-10$ for $6M_{\odot}-8M_{\odot}$ stars) and least pronounced for the least massive stars (with a factor of $1-2$ for $3M_{\odot}-4M_{\odot}$ stars).  Because of the rarity of Milky Way stars with masses greater than or equal to $3M_{\odot}$, we do not expect the increased rate of ejections during the WD phase to provide a major source for the current free-floating planet population.  However, if future work can show that the potentially complex orbital dynamics accompanying  $M\lesssim 1 M_{\odot}$ stellar evolution yields a planetary ejection rate around WDs that is several factors higher than the MS rate, then the Galaxy's free-floating planet population may be a strong function of time.

\section{Conclusions}

\label{sec:conclusions}

We have carried out numerical integrations of systems of three giant planets orbiting $3-8\mathrm{\,M}_\odot$ stars up to a total system age of 5\,Gyr, covering all the star's evolutionary stages from the beginning of the MS and for several Gyr of WD cooling. We have studied systems of planets on wide orbits, initially beyond 10\,au, which are unaffected by stellar evolutionary effects such as tides, feeling only the changing stellar mass. By following systems through the entirety of the star's lifetime, we ensure that the systems we simulate on the post-MS have survived MS evolution.

In common with previous scattering studies, we find that on the MS stability is dependent on the planets' initial spacing, with planets separated by less than 9 (single-planet) Hill radii prone to disruption. The outcome of these instabilities is the ejection of a planet or the collision of one planet with another or with the star. In some cases two planets are lost from the systems, leaving only one survivor. Loss of all planets on the MS does not occur in our simulations.

When the host star evolves off the MS, the increase in stellar radius can engulf planets in systems that experienced instability on the MS, where these planets are left on eccentric orbits with pericentres close to the star. Instability can also leave a surviving planet with a large apocentre, in which case the planet can become unbound as a result of the star's mass loss. In a very few cases one of these effects succeeds in removing the system's third and final planet, leaving the star totally denuded of its planetary entourage. This happens but rarely however, in about 1\% of all our integrations. This implies that the fraction of MS stars hosting planets on orbits of $\sim10$ to a few hundreds of au should be almost identical to the fraction of their WD descendants hosting planets at similar radii.

Few hitherto-stable systems manifest instability between the end of the MS and the start of the WD stage, due to the relatively short duration of this evolutionary phase. Nor does the mass loss at the end of the AGB immediately trigger destabilisation of previously-stable systems. Instead, instability around WDs becomes manifest typically at cooling ages of several hundred Myr. Instability in WD systems strikes both systems that survived previous evolution unscathed, and systems that previously lost a planet and have two remaining on eccentric orbits. In the latter case instability occurs at slightly later cooling ages, with a median of $10^{8.4}$\,yr as opposed to $10^{7.9}$\,yr in our $1\mathrm{\,M_J}$ integrations.

The overwhelming outcome of WD instability is the ejection of at least one planet. Collisions between planets or between a planet and a star are relatively rare. We find that very few planets are scattered onto orbits with small pericentres, where they may then become tidally circularised and discovered as WD-transiting planets. Combining our simulation results with planet detection rates around MS stars, we expect $\lesssim1\%$ of WDs to host close-in planets.

Similarly, we expect that the pollution of WDs following instability in systems of giant planets, whether directly through the accretion of a planet itself or indirectly through the destabilisation of small bodies following planetary instability, would occur in $\lesssim7\%$ of WDs, insufficient to explain observed pollution rates.

Although we restricted our integrations to a particular range of parameters, we have made predictions for behaviour for other parameter values. In particular, we expect the number of systems destabilised around low-mass WDs with progenitors $<3\mathrm{\,M}_\odot$ to be much lower, as a consequence of the longer MS lifetimes and particularly the smaller amount of mass loss.

\section*{Acknowledgements}

AJM and EV are supported by Spanish grant AYA 2010/20630. EV also acknowledges the support of the Marie Curie grant FP7-People-RG268111. Additionally, the research leading to these results has received funding from the European Research Council under the European Union's Seventh Framework Programme (FP/2007-2013) / ERC Grant Agreement n. 320964 (WDTracer). We thank the anonymous referee for inspiring a number of points of clarification in the paper. We thank Boris G\"ansicke and Amy Bonsor for comments on the manuscript.

\bibliographystyle{mn2e-long}
\bibliography{bibliography}

\bsp

\label{lastpage}

\end{document}